\documentstyle[amssymb,aps,preprint]{revtex}

\begin{document}
\author{M. Fuchs, W. G\"{o}tze, and M.R. Mayr}
\address{Physik--Department, Technische Universit\"{a}t M\"{u}nchen, 85747 Garching}
\title{Asymptotic laws for tagged-particle motion in glassy systems}
\date{Phys. Rev. E, in print}
\maketitle

\begin{abstract}
Within the mode--coupling theory for structural relaxation in simple systems
the asymptotic laws and their leading--asymptotic correction formulas are
derived for the motion of a tagged particle near a glass--transition
singularity. These analytic results are compared with numerical ones of the
equations of motion evaluated for a tagged hard sphere moving in a
hard--sphere system. It is found that the long--time part of the two--step
relaxation process for the mean--squared displacement can be characterized
by the $\alpha $--relaxation--scaling law and von Schweidler's power--law
decay while the critical--decay regime is dominated by the corrections to
the leading power--law behavior. For parameters of interest for the
interpretations of experimental data, the corrections to the leading
asymptotic laws for the non--Gaussian parameter are found to be so large
that the leading asymptotic results are altered qualitatively by the
corrections. Results for the non-Gaussian parameter are shown to follow
qualitatively the findings reported in the molecular--dynamics--simulations
work by Kob and Andersen [Phys. Rev. E {\bf 51}, 4626 (1995)].
\end{abstract}

\bigskip

PACS numbers: 64.70 Pf, 61.20 Lc

\section{Introduction}

The mode--coupling theory (MCT) for the structural relaxation of glassy
liquids \cite{Leutheusser84,Bengtzelius84} has provided explanations for
long--known phenomena as for example the stretching of the $\alpha $
process, the time--temperature superposition principle, and the von
Schweidler law. It has also predicted new phenomena such as a square--root
singularity in the temperature dependence of the Debye--Waller factor and
the critical decay. For details the reader is referred to Ref. \cite
{Goetze91b,Goetze92} and the papers quoted there. The universal features of
the MCT arise as leading--order results near a bifurcation singularity,
which was identified as an ideal glass transition---in reality smeared out
due to additional relaxation processes. These theoretical findings also
provided some motivation to a number of experiments focusing on the
evolution of structural relaxation in the vicinity of the predicted
glass--transition singularity (see Ref. \cite
{Mezei91,Li92,Bartsch92,Megen93,Megen94b,Baschnagel94,Kob95,Yang96,Ma96,Gallo96,Lunkenheimer97b,Cummins97,Toelle97}
and the papers quoted there). In turn these experiments have sparked off new
theoretical developments such as the derivation and discussion of
next--to--leading--order asymptotic results in the vicinity of the
bifurcation singularity \cite{Franosch97} and the extension of the theory to
liquids consisting of or containing non--spherical molecules \cite
{Schilling97,Franosch97c}. Using the coherent intermediate scattering
function of a hard--sphere system (HSS) as an example, the preceding work 
\cite{Franosch97} illustrates how asymptotic corrections explain the
deviation from the leading--order results and shows how general properties
of these corrections may be included into a quantitative analysis of
experimental data. Due to the presence of additional amplitudes and a new
second--order scaling function, many features of these corrections are
different for different observables, used for the description of the glassy
dynamics. Therefore we intend to extend the previous work to a discussion of
the conceptually simplest functions characterizing the dynamics, namely the
density correlator of a tagged particle and its special limits, the
mean--squared displacement and the non--Gaussian parameter.

The tagged--particle density correlators can be measured by a variety of
techniques. Neutron scattering from molecular liquids with an incoherent
scattering cross--section such as orthoterphenyl \cite{Toelle97,Petry91}
probes its spectrum. For a glassy colloidal suspension the correlator has
recently been measured by dynamic light scattering \cite{Megen98}. From
measurements of the incoherent intermediate scattering function for small
wave vectors one can extract the mean--squared displacement \cite
{Megen98,Zorn97,Kanaya97} and, in principle, also information about the
non--Gaussian parameter. In computer simulations the tagged--particle
correlator is a preferred quantity---compared to the coherent density
fluctuation function---due to statistical advantages. The mean--squared
displacement and the non--Gaussian parameter can also be directly extracted
from computer simulation data\cite{Kob95,Sciortino96}. The reported
simulation results for the non--Gaussian parameter challenged the universal
features predicted by MCT. Within the accessible simulation--time window the
non--Gaussian parameter showed no signs of the predictions of a two--step
relaxation process nor of the time--temperature superposition principle,
even though other dynamical quantities such as the intermediate scattering
function, self--intermediate scattering function and the mean--squared
displacement fitted into the picture for glassy dynamics drawn by the MCT.

To proceed, equations of motion will be derived for the mean--squared
displacement and the non--Gaussian parameter. The relaxation kernels in
those equations are given by mode--coupling functionals, which require the
correlators of the density and the tagged--particle density as input. For
the latter the previously--derived MCT equations \cite{Bengtzelius84} are
used. Then an asymptotic expansion of the solutions will be carried out. All
equations will be solved and studied in quantitative detail for the model
system of a tagged hard sphere immersed in a HSS. The HSS has already been
the subject of a number of theoretical investigations as can be inferred
from Refs. \cite{Franosch97,Fuchs95} and the papers quoted there. Its
importance stems in part from the fact that it is the simplest system for
which a glass transition has been detected experimentally \cite
{Megen93,Megen94b,Megen91}, thus providing the archetype for quantitative
tests of the theory. 
Comprehensive comparisons between MCT results for the evolution of structural
relaxation and the corresponding experimental findings obtained by dynamic
light scattering for hard sphere colloids have been published in Refs. 
\cite{Megen93,Megen94b,Megen91n,Megen93n}.In the data analysis enter two
numbers as fit parameters: a time scale $t_0$, given by the viscosity of the
solvent, and the critical packing fraction $\varphi_c$. The predicted value
for $\varphi_c$ \cite{Bengtzelius84} differs from the the experimental value by
about $12\%$. Here as in any other test of a singularity theory, data have been
studied as a function of the distance 
$\varepsilon=(\varphi-\varphi_c)/\varphi_c$ from the position $\varphi_c$ of
the singularity. The cited tests have been summarized as follows
\cite{Megen94n,Megen95n}: the leading asymptotic MCT results account for the
data within the experimental uncertainties for the liquid states
($\varepsilon<0$) and also for the glass states ($\varepsilon>0$). This does
not only hold for universal formulae like the scaling laws. It holds also for
the numerical values predicted for the exponents, the master functions entering
the scaling laws, and for the various wave--vector--dependent
amplitudes. Therefore it seems justified to illustrate our theory for the
tagged particle motion in a HSS. The cited and tested results for the density
dynamics will enter as input taken from the preceding work \cite{Franosch97}. 
But the HSS is also relevant because its dynamics shares
many features with other simple systems as will again be demonstrated in
this article.

The paper is arranged as follows. After briefly reviewing the basics of the
MCT glass transition (Sect. \ref{GTHSM}) we start by discussing the
properties of the tagged--particle density correlator in Sect. \ref{SISF}.
The next Section covers the mean--squared displacement and its asymptotic
behavior. The quality of the Gaussian approximation for the
self--intermediate scattering function of the HSS is assessed (Sect. \ref{GA}%
) before turning to the non--Gaussian parameter in Sect. \ref{NGP}. In the
conclusion we will summarize our results and suggest how some recent
experiments and simulations on single--particle quantities can be
interpreted.

\section{The ideal MCT glass transition\label{GTHSM}}

The idealized MCT deals with the structural dynamics of simple liquids,
consisting of $N$ particles at positions $\vec{r}_{j}$, $j=1,\ldots ,N$. A
self--consistent treatment of the cage effect, which is thought to be the
reason for the glass transition, leads to a closed set of
integro--differential equations for the density correlator $\Phi _{q}\left(
t\right) =\left\langle \rho _{\vec{q}}^{*}(0)\rho _{\vec{q}}(t)\right\rangle
/S_{q}$, i.e., the autocorrelation function of the density fluctuation $\rho
_{\vec{q}}(t)=1/\sqrt{N}\sum_{j}\exp \left[ i\vec{q}\vec{r}_{j}\left(
t\right) \right] $ for wave vector $\vec{q}$, normalized by the static
structure factor $S_{q}=\left\langle \rho _{\vec{q}}^{*}(0)\rho _{\vec{q}%
}(0)\right\rangle $; $\left\langle {}\right\rangle $ signifies an average
with respect to the canonical ensemble and $q=\left| \vec{q}\right| $
abbreviates a wave vector modulus. Specializing to a colloid model, the MCT
equations of motion read 
\begin{equation}
\tau _{q}\dot{\Phi}_{q}\left( t\right) +\Phi _{q}\left( t\right)
+\int\limits_{0}^{t}m_{q}\left( t-t^{\prime }\right) \dot{\Phi}_{q}\left(
t^{\prime }\right) dt^{\prime }=0.  \label{MCE_ISF}
\end{equation}
They are specified by the times $\tau _{q}=S_{q}/\left( D_{0}q^{2}\right) $
with $D_{0}$ denoting the single particle diffusion coefficient. The memory
kernel in this equation is given as a mode--coupling functional $m_{q}\left(
t\right) ={\cal F}_{q}\left( \Phi \left( t\right) \right) $, where ${\cal F}%
_{q}$ is determined by the structure factor $S_{q}$ \cite{Bengtzelius84},
which depends smoothly on external control parameters such as the density $n$%
.

At a critical density $n_{c}$ the above equations exhibit a bifurcation in
the long--time limits $f_{q}=\lim_{t\rightarrow \infty }\Phi _{q}\left(
t\right) $ of the density correlators. The bifurcation singularity can be
identified with a liquid--glass transition: At $n_{c}$, the Debye--Waller
factor $f_{q}$, also called form factor, jumps from zero to the critical
from factor $f_{q}^{c}>0$. The instability of the glass for $n\rightarrow
n_{c}+$ is reflected by a square--root singularity of the Debye--Waller
factor: $f_{q}=f_{q}^{c}+h_{q}\sqrt{\sigma /\left( 1-\lambda \right) }+{\cal %
O}\left( \sigma \right) $. Here, $\sigma =C\varepsilon $ is called the
separation parameter, $0.5\leq \lambda <1$ is called exponent parameter, and 
$h_{q}=\left( 1-f_{q}^{c}\right) ^{2}e_{q}>0$ denotes the critical
amplitude. The quantities $e_{q}$, $\lambda $, and the constant $C>0$,
relating $\sigma $ and the reduced density $\varepsilon =\left(
n-n_{c}\right) /n_{c}$, can be calculated by straight--forward formulae from
the functional ${\cal F}$ \cite{Goetze91b}.

Via a simple transcendental equation the exponent parameter $\lambda $
determines two anomalous exponents: the critical exponent $a$, $0<a<0.5$,
and the von Schweidler exponent $b$, $0<b\leq 1$. Furthermore $\lambda $
fixes a constant $B>0$. There appears a single time $t_{0}$, specifying the
scale for the transient dynamics. The bifurcation dynamics is then ruled by
two critical time scales, denoted by $t_{\sigma }$ and $t_{\sigma }^{\prime
} $: 
\begin{equation}
t_{\sigma }=t_{0}/\left| \sigma \right| ^{\delta },t_{\sigma }^{\prime
}=t_{0}B^{-1/b}/\left| \sigma \right| ^{\gamma },\delta =\frac{1}{2a},\gamma
=\frac{1}{2a}+\frac{1}{2b}.  \label{timescales}
\end{equation}
The mathematical relevance of these concepts is evident from the following
limit results \cite{Goetze91b}. First, $\lim_{\hat{t}\rightarrow
0}\lim_{\sigma \rightarrow 0}\left\{ \left[ \Phi _{q}\left( \hat{t}t_{\sigma
}\right) -f_{q}^{c}\right] /\sqrt{\left| \sigma \right| }\right\} \hat{t}%
^{a}=h_{q}$. Thus, near the transition the plateau value $f_{q}^{c}$ is
approached from above according to a power law, called critical decay 
\begin{equation}
\Phi _{q}\left( t\right) -f_{q}^{c}\sim h_{q}\left( t_{0}/t\right)
^{a},t_{0}\ll t\ll t_{\sigma }.  \label{critlaw}
\end{equation}
Second, $\lim_{\tilde{t}\rightarrow 0}\lim_{\sigma \rightarrow 0-}\left[
\Phi _{q}\left( \tilde{t}t_{\sigma }^{\prime }\right) -f_{q}^{c}\right] /%
\tilde{t}^{b}=-h_{q}=\lim_{\hat{t}\rightarrow \infty }\lim_{\sigma
\rightarrow 0-}\left\{ \left[ \Phi _{q}\left( \hat{t}t_{\sigma }\right)
-f_{q}^{c}\right] /\sqrt{\left| \sigma \right| }\right\} /\hat{t}^{b}$. Near
the transition the plateau value $f_{q}^{c}$ is left in the liquid state
according to a power law, called von Schweidler decay 
\begin{equation}
\Phi _{q}\left( t\right) -f_{q}^{c}\sim -h_{q}\left( t/t_{\sigma }^{\prime
}\right) ^{b},t_{\sigma }\ll t\ll t_{\sigma }^{\prime }.  \label{vSchwlaw}
\end{equation}

There are two dynamical scaling laws describing the bifurcation dynamics
near the transition in a leading--order asymptotic limit. The first one
describes the dynamics on a scale $t_{\sigma }$: $\Phi _{q}\left( t\right)
-f_{q}^{c}\sim h_{q}G\left( t\right) $. Here $G\left( t\right) =\left(
t_{0}/t\right) ^{a}$ for $\sigma =0$ and 
\begin{equation}
G\left( t\right) =\sqrt{\left| \sigma \right| }g_{\pm }\left( t/t_{\sigma
}\right) ,\sigma \gtrless 0.  \label{beta_scaling}
\end{equation}
The $\sigma $--independent master functions $g_{\pm }$ are determined by $%
\lambda $. They describe the crossover from the critical decay $g_{\pm
}\left( \hat{t}\ll 1\right) =1/\hat{t}^{a}$ to arrest in the glass, $%
g_{+}\left( \hat{t}\gg 1\right) =1/\sqrt{1-\lambda }$, or to von
Schweidler's decay in the liquid, $g_{-}\left( \hat{t}\gg 1\right) =-B\hat{t}%
^{b}+B_{1}/\left( B\hat{t}^{b}\right) $. The result is obtained by solving
the MCT equations using $\left| \Phi _{q}\left( t\right) -f_{q}^{c}\right| $
as small parameter. This is equivalent to writing $t=\hat{t}t_{\sigma }$ and
then expanding in powers of the small parameter $\sqrt{\left| \sigma \right| 
}$. Pushing the expansion to next--to--leading order extends the first
scaling--law result to 
\begin{equation}
\Phi _{q}\left( t\right) =f_{q}^{c}+h_{q}G\left( t\right) +h_{q}\left[
H\left( t\right) +K_{q}G\left( t\right) ^{2}+\sigma \hat{K}_{q}\right] .
\label{AE_ISF}
\end{equation}
Here two further amplitudes $K_{q}$ and $\hat{K}_{q}$ and a new function $%
H\left( t\right) $ appear. The latter obeys a scaling law: $H\left( t\right)
=\kappa \left( a\right) \left( t_{0}/t\right) ^{2a}$ for $\sigma =0$ and $%
H\left( t\right) =\left| \sigma \right| h_{\pm }\left( t/t_{\sigma }\right) $
for $\sigma \gtrless 0$. All these quantities can be evaluated from the
mode--coupling functional ${\cal F}_{q}$ as is comprehensively explained in
Ref. \cite{Franosch97}, where we changed notation 
$\bar{\bar{K}}$%
to $\hat{K}_{q}$. The domain of applicability of the expansion (\ref{AE_ISF}%
) is called $\beta $--relaxation window, and correspondingly $G$ and $%
t_{\sigma }$ are respectively referred to as $\beta $ correlator and $\beta $%
--relaxation time scale.

The second scaling law describes the liquid dynamics on scale $t_{\sigma
}^{\prime }$. It deals with the decay of the correlator from the plateau to
zero and it is based on the relation $\lim_{\sigma \rightarrow 0-}\Phi
_{q}\left( \tilde{t}t_{\sigma }^{\prime }\right) =\tilde{\Phi}_{q}\left( 
\tilde{t}\right) $. Here the $\sigma $--independent functions $\tilde{\Phi}%
_{q}\left( \tilde{t}\right) $ are the solution of the implicit functional
equations 
\begin{equation}
\int\limits_{0}^{\tilde{t}}\left[ \tilde{m}_{q}\left( t^{\prime }\right) -%
\tilde{\Phi}_q\left( t^{\prime }\right) \right] dt^{\prime }=\int\limits_{0}^{%
\tilde{t}}\tilde{m}_{q}\left( \tilde{t}-t^{\prime }\right) \tilde{\Phi}_q%
\left( t^{\prime }\right) dt^{\prime }.  \label{alpha_ISF}
\end{equation}
The kernel $\tilde{m}_{q}\left( \tilde{t}\right) ={\cal F}_{q}^{c}\left( 
\tilde{\Phi}\left( \tilde{t}\right) \right) $ is the mode--coupling
functional at the transition point and the equation is to be solved with von
Schweidler's law as initial condition. Near the transition within the window 
$t_{\sigma }\ll t$ this implies in leading order 
\begin{equation}
\Phi _{q}\left( t\right) =\tilde{\Phi}_{q}\left( t/t_{\sigma }^{\prime
}\right) .  \label{TDSP}
\end{equation}
This window is referred to as the $\alpha $ regime, $t_{\sigma }^{\prime }$
is called $\alpha $--relaxation time scale, and Eq. (\ref{TDSP}) is the
superposition principle for the $\alpha $ process.

Every theory dealing with the dynamics of a variable $A$ as a probe of
structural relaxation in glassy systems relies on an understanding of the
dynamics of density rearrangements. Hence a microscopic theory will require
results for $\Phi _{q}\left( t\right) $ as input. For the following
discussion it is therefore necessary to appreciate the concepts and results
formulated above. They are explained and illustrated in Ref. \cite
{Franosch97} for the HSS. The quantitative examples of the following
discussion shall also be done for that model. We follow the previous
conventions by using the particle diameter $d$ as unit of length, $d=1$,
choosing the time unit so that the short time diffusivity $D_{0}=1/160$, and
using the packing fraction $\varphi =\pi nd^{3}/6$ as the control parameter.
In this case one gets: $\varphi _{c}=0.516$, $C=1.54$, $\lambda =0.735$, $%
a=0.312$, $b=0.583$, $B=0.836$, $B_{1}=0.431$ ,$t_{0}=0.425$.

\section{The tagged--particle density correlator\label{SISF}}

\subsection{Equation of motion}

The dynamics of a tagged particle is explored by its density correlator $%
\Phi _{q}^{s}\left( t\right) =\left\langle \rho _{\vec{q}}^{s*}\left(
0\right) \rho _{\vec{q}}^{s}\left( t\right) \right\rangle $, where $\rho _{%
\vec{q}}^{s}\left( t\right) =\exp \left[ i\vec{q}\vec{r}\left( t\right)
\right] $; $\vec{r}$ denotes the position of the particle. For a system
driven by Brownian dynamics the MCT equation of motion reads 
\begin{equation}
\tau _{q}^{s}\dot{\Phi}_{q}^{s}\left( t\right) +\Phi _{q}^{s}\left( t\right)
+\int\limits_{0}^{t}m_{q}^{s}\left( t-t^{\prime }\right) \dot{\Phi}%
_{q}^{s}\left( t^{\prime }\right) dt^{\prime }=0  \label{MCE_SISF}
\end{equation}
with $\tau _{q}^{s}=1/\left( D_{0}^{s}q^{2}\right) $. Here $D_{0}^{s}$ is
the short--time diffusion coefficient of the immersed particle. The memory
kernel $m_{q}^{s}\left( t\right) ={\cal F}_{q}^{s}\left( \Phi \left(
t\right) ,\Phi ^{s}\left( t\right) \right) $ is expressed through the
mode--coupling functional ${\cal F}_{q}^{s}$, where not only $\Phi
_{q}^{s}\left( t\right) $ but also $\Phi _{q}\left( t\right) $ enter: 
\begin{equation}
{\cal F}_{q}^{s}\left( f,f^{s}\right) =\frac{1}{\left( 2\pi \right) ^{3}}%
\int nS_{k}c_{k}^{s2}\left( \frac{\vec{q}\vec{k}}{q^{2}}\right) ^{2}f_{k}f_{|%
\vec{q}-\vec{k}|}^{s}\,d^{3}k.  \label{MCF_SISF}
\end{equation}
Here $c_{q}^{s}=\left\langle \rho _{\vec{q}}^{s*}(0)\rho _{\vec{q}%
}(0)\right\rangle /\left( nS_{q}\right) $ denotes the single--particle
direct correlation function \cite{Bengtzelius84}.

To solve Eqs. (\ref{MCE_SISF}) and (\ref{MCF_SISF}) numerically we proceed
as in Ref. \cite{Franosch97}: The mode--coupling functional is rewritten in
bipolar coordinates by change of variable and the two remaining integrals
are approximated by Riemann sums 
\begin{equation}
{\cal F}_{q}^{s}\left( f,f^{s}\right) =n\Delta ^{3}/\left( 16d^{3}\pi
^{2}\right) \sum_{\hat{k}=\frac{1}{2}}^{99\frac{1}{2}}\sum_{\hat{p}=\frac{1}{%
2}}^{99\frac{1}{2}}{}^{\prime }S_{k}\left( \hat{k}\hat{p}/\hat{q}^{5}\right)
\left( \hat{k}^{2}+\hat{q}^{2}-\hat{p}^{2}\right)
^{2}c_{k}^{s2}f_{k}f_{p}^{s}.  \label{DMCF_SISF}
\end{equation}
The prime indicates that the sum runs only over those values $\hat{p}$ for
which $\hat{k}$, $\hat{p}$, and $\hat{q}$ obey the triangle inequality.
Thereby, we again get a precisely defined tagged--hard--sphere--particle
model described by $100$ coupled integro--differential equations for $\Phi
_{q}^{s}\left( t\right) $ on a grid of $100$ equally spaced wave numbers $%
q=\Delta \hat{q}$ with $\hat{q}=\frac{1}{2},\ldots ,99\frac{1}{2}$ and step
size $\Delta =0.4$.

The substitution of Eq. (\ref{DMCF_SISF}) for the mode--coupling functional (%
\ref{MCF_SISF}) might seem rather crude. However we have checked that the
results seldom differ by more than a few percent from those obtained with $%
300$ or $900$ grid points for the Riemann sum. Only at small wave numbers
qualitative changes occur due to bad angle resolution; but these do not
influence other wave vectors because the major contributions to the integral
in Eq. (\ref{MCF_SISF}) come from the wave vectors near the
structure--factor--peak position.

In our application to the HSS the static structure factor $S_{q}$ and the
direct correlation function $c_{q}^{s}$ are calculated in the Percus--Yevick
approximation \cite{Lebowitz63}. Assuming short--time diffusion according to
Stokes' law, which entails $D_{0}^{s}/D_{0}=d/d^{s}$, only two control
parameters remain: the packing fraction $\varphi $ of the host particles and
the diameter $d^{s}$ of the tagged particle. All results for the coherent
density correlators $\Phi _{q}\left( t\right) $ of the HSS, which will be
needed in the following, will be taken from the solution of Eq. (\ref
{MCE_ISF}) discussed in Ref. \cite{Franosch97}.

\subsection{The critical Lamb--M\"{o}ssbauer factor\label{ASB_SISF}}

The trapping of a tagged--particle by its surrounding host particles
manifests itself by a nonvanishing Lamb--M\"{o}ssbauer factor $%
f_{q}^{s}=\lim_{t\rightarrow \infty }\Phi _{q}^{s}\left( t\right) $. From
Eq. (\ref{MCE_SISF}) one finds \cite{Bengtzelius84} that $f_{q}^{s}$ is a
solution of 
\begin{equation}
\frac{f_{q}^{s}}{1-f_{q}^{s}}={\cal F}_{q}^{s}\left( f,f^{s}\right) .
\label{LMF}
\end{equation}
It is distinguished from other solutions of Eq. (\ref{LMF}), say $\tilde{f}%
_{q}^{s}$, by the maximum property, $f_{q}^{s}\geq \tilde{f}_{q}^{s}$ \cite
{Goetze91b}. It can be found by the iteration: $f_{q}^{s}=\lim_{n\rightarrow
\infty }f_{q}^{s(n)}$, where $f_{q}^{s(n+1)}/\left( 1-f_{q}^{s(n+1)}\right) =%
{\cal F}_{q}^{s}\left( f,f^{s(n)}\right) $, $f_{q}^{s(0)}=1$\cite{Goetze95b}%
. Obviously $f\equiv 0$ implies $f^{s}\equiv 0$, but not vice versa. The
spatial Fourier transform $f^{s}\left( r\right) $ of the Lamb--M\"{o}ssbauer
factor is the probability of finding the particle for $t=\infty $ at
distance $r$ from where it started for $t=0$. Therefore, a particle cannot
be trapped as long as the host particles are in a liquid state. But even
when the host particles are in a glass state the tagged particle may still
be able to diffuse in the arrested structure. The various scenarios arising
from Eqs. (\ref{LMF}) and (\ref{MCF_SISF}) have been discussed before
(compare Ref. \cite{Franosch94} and the papers cited there).

For a tagged hard sphere of diameter $d^{s}=1.0$ the Lamb--M\"{o}ssbauer
factor jumps from $0$ to the critical value $f_{q}^{sc}>0$, which is shown
in Fig. \ref{CritAmpR1}, when the packing fraction $\varphi $ passes through
the glass--transition singularity $\varphi _{c}$. For $d^{s}=0.6$ the
particle still gets trapped at $\varphi _{c}$ but the $f_{q}^{sc}$--versus--$%
q$ graph is narrower than before (see Fig.\ref{CritAmpR06}). This means that
the centre of the smaller particle can explore a larger volume. Reducing the
particle size $d^{s}$ further, the system passes through a percolation
threshold at some critical diameter $d^{sc}$. In the following we will only
be interested in tagged particles with diameters $d^{s}>d^{sc}$, for which
the Lamb--M\"{o}ssbauer factor shows the generic fold bifurcation.

\subsection{Asymptotic laws}

\subsubsection{$\beta $ relaxation\label{BETA_SISF}}

Close to the bifurcation the correlators $\Phi _{q}^{s}\left( t\right) $
show a characteristic two--step relaxation process, as can be seen in Fig. 
\ref{phis of t}. In Ref. \cite{Goetze92} it was discussed and in Ref. \cite
{Franosch98} it was demonstrated for relevant examples that the long--time
behavior of correlators, which are the solutions of MCT equations such as (%
\ref{MCE_ISF}) and (\ref{MCE_SISF}), is determined by 
\begin{equation}
\frac{\Phi _{q}^{s}(s)}{1-s\Phi _{q}^{s}(s)}={\cal L}[{\cal F}_{q}^{s}(\Phi
(t),\Phi ^{s}(t))](s).  \label{SSR}
\end{equation}
Here $\Phi _{q}^{s}(s)$ denotes the Laplace transform $\Phi \left( s\right) =%
{\cal L}\left[ \Phi \left( t\right) \right] \left( s\right)
:=\int_{0}^{\infty }e^{-st}\Phi \left( t\right) dt$ of $\Phi _{q}^{s}\left(
t\right) $. These equations depend only on equilibrium quantities via the
static structure factor. They are therefore independent of the microscopic
dynamics and any additional regular contributions to the kernels $%
m_{q}\left( t\right) $ and $m_{q}^{s}\left( t\right) $. Consequently, it
makes no difference for the asymptotic results discussed throughout this
paper that we chose Brownian instead of Newtonian dynamics and omitted
regular contributions to the kernels.

Inserting the asymptotic expansion (\ref{AE_ISF}) into Eq. (\ref{SSR}),
using Taylor's theorem on the mode--coupling functional ${\cal F}_{q}^{s}$,
and collecting like--order terms one obtains in leading and
next--to--leading order 
\begin{equation}
\Phi _{q}^{s}\left( t\right) =f_{q}^{sc}+h_{q}^{s}G\left( t\right)
+h_{q}^{s}\left[ H\left( t\right) +K_{q}^{s}G\left( t\right) ^{2}+\sigma 
\hat{K}_{q}^{s}\right] .  \label{AE_SISF}
\end{equation}
Here the critical amplitude $h_{q}^{s}=\left( 1-f_{q}^{sc}\right)
^{2}e_{k}^{s}$ for the tagged particle correlator and the correction
amplitudes $K_{q}^{s}$ and $\hat{K}_{q}^{s}$ are to be calculated from 
\begin{mathletters}
\label{AMP_SISF}
\begin{eqnarray}
\sum_{k}\left( \delta _{qk}-C_{qk}^{sc}\right) e_{k}^{s}
&=&\sum_{k}C_{q,k}^{sc}e_{k}, \\
\sum_{k}\left( \delta _{qk}-C_{qk}^{sc}\right) e_{k}^{s}K_{k}^{s} &=&-\left(
1-f_{q}^{sc}\right) e_{q}^{s2}\lambda +\sum_{k}C_{q,k}^{sc}e_{k}K_{k} 
\nonumber \\
&&+\sum_{p,k}C_{qpk}^{sc}e_{p}^{s}e_{k}^{s}+%
\sum_{p,k}C_{q,pk}^{sc}e_{p}e_{k}+\sum_{p,k}C_{qp,k}^{sc}e_{p}^{s}e_{k}, \\
\sum_{k}\left( \delta _{qk}-C_{qk}^{sc}\right) e_{k}^{s}\hat{K}_{k}^{s}
&=&-\left( 1-f_{q}^{sc}\right) e_{q}^{s2}+\delta
C_{q}^{s}/C+\sum_{k}C_{q,k}^{sc}e_{k}\hat{K}_{k}.
\end{eqnarray}
Taylor coefficients of the mode--coupling functional are introduced by 
\end{mathletters}
\begin{eqnarray}
C_{qp_{1}\ldots p_{n},k_{1}\ldots k_{m}}^{s} &=&C_{qp_{1}\ldots
p_{n},k_{1}\ldots k_{m}}^{sc}+\delta C_{qp_{1}\ldots p_{n},k_{1}\ldots
k_{m}}^{s}\varepsilon +{\cal O}\left( \varepsilon ^{2}\right)  \nonumber \\
&=&\frac{1}{n!}\frac{1}{m!}\frac{\partial ^{n}\partial ^{m}{\cal F}%
_{q}^{s}\left( f^{c},f^{sc}\right) }{\partial f_{p_{1}}\ldots \partial
f_{p_{n}}\partial f_{k_{1}}\ldots \partial f_{k_{m}}}\times  \nonumber \\
&&\left( 1-f_{p_{1}}^{sc}\right) ^{2}\ldots \left( 1-f_{p_{n}}^{sc}\right)
^{2}\left( 1-f_{k_{1}}^{c}\right) ^{2}\ldots \left( 1-f_{k_{m}}^{c}\right)
^{2}.  \label{DefCs}
\end{eqnarray}
For the derivation of these general formulae we have made no use of the
linearity of the mode--coupling functional in $f$ and $f^{s}$. Figures \ref
{CritAmpR1} and \ref{CritAmpR06} exhibit the HSS results for the amplitudes
for the two diameters $d^{s}=1.0$ and $d^{s}=0.6$, respectively. Following
Ref. \cite{Franosch97} we show 
\begin{equation}
\bar{K}_{q}^{s}=\hat{K}_{q}^{s}\sqrt{1-\lambda }+K_{q}^{s}/\sqrt{1-\lambda }%
-\kappa +h_{+}\left( \infty \right) \sqrt{1-\lambda }  \label{Kbar}
\end{equation}
instead of $\hat{K}_{q}^{s}$, where $\kappa =0.961$ and $h_{+}\left( \infty
\right) =-2.21$ for the HSS.

Equation (\ref{AE_SISF}) for the tagged--particle correlator is formally
identical to the above expansion (\ref{AE_ISF}) for the coherent density
correlator. This reflects the universality of the asymptotic features of the
MCT. Within the $\beta $--relaxation window the dynamics is fixed by the $%
\beta $ correlator $G\left( t\right) $ in leading order and by $G$ and $H$
in next--to--leading order in an asymptotic expansion. Correlators referring
to different probing variables merely differ in the various amplitudes such
as ($f_{q}^{c}$,$f_{q}^{sc}$), ($h_{q}$,$h_{q}^{s}$), etc. These amplitudes
do not depend on time nor on control parameters. However, there are rather
important differences for the procedure to determine the amplitudes
reflecting the fact that the bifurcation in the tagged--particle dynamics is
brought about by the bifurcation in the host--particle dynamics. The
amplitudes $e_{q}$, $K_{q}$ and $\hat{K}_{q}$ for the cage--forming
particles are the solutions of singular matrix equations and the solubility
conditions in second and third order of the asymptotic expansion determine
the scaling functions $G$ and $H$ up to a single time scale $t_{0}$, which
has to be matched at the critical point \cite{Franosch97}. By contrast, the
positive matrix $C_{qk}^{sc}$ has a spectral radius smaller than unity \cite
{Goetze95b}. Hence, the matrix $\delta _{qk}-C_{qk}^{sc}$ is regular and the
amplitudes are uniquely determined by solving the linear equations (\ref
{AMP_SISF}). The scaling functions $G$ and $H$ are solely determined by the
properties of the host particles. The interaction potential of the tagged
particle enters the Lamb--M\"{o}ssbauer factor and the amplitudes via the
direct correlation function $c_{q}^{s}$, which is hidden in the Taylor
coefficients (\ref{DefCs}). No further time scale needs to be fixed. Even a
change of the microscopic time scales $\tau _{q}^{s}$ in Eq. (\ref{MCE_SISF}%
) would not affect the time scale of the asymptotic behavior. One can always
go so close to the critical point that Eq. (\ref{AE_SISF}) holds within a
prescribed error margin.

The first two terms on the right--hand side of Eq. (\ref{AE_SISF}) are known
as the factorization theorem: If the density correlators are rescaled
accordingly the curves collapse onto the $\beta $ correlator $G$ in an
intermediate time window, the so--called $\beta $--relaxation region, as is
demonstrated in Fig. \ref{red_theo} for four wave numbers. Obviously, the $%
\beta $--relaxation region depends on the wave number (upper panel) or---put
more generally---on the quantity under consideration (lower panel). This is
accounted for by the two $q$--dependent amplitudes $K_{q}^{s}$ and $\hat{K}%
_{q}^{s}$ occurring in the leading correction to the factorization property.
The $\beta $ window expands as the system moves closer to the critical point
as is demonstrated by a comparison of $\Phi _{q}^{s}\left( t\right) $ for $%
q_{2}=10.6$ for $n=9$ and $n=10$ in Fig. \ref{red_theo}. More can be said if
we consider that the dominant corrections are the short--time corrections to
the critical law (\ref{critlaw}), and the long--time corrections to the von
Schweidler law (\ref{vSchwlaw}) or the corrections to the
Lamb--M\"{o}ssbauer factor $f_{q}^{s}$ on the liquid ($\varepsilon <0$) or
glass ($\varepsilon >0$) side, respectively. The short--time expansion of
the right--hand side of Eq. (\ref{AE_SISF}) is given by $%
f_{q}^{sc}+h_{q}^{s}\left( t/t_{0}\right) ^{-a}\left\{ 1+\left[ \kappa
\left( a\right) +K_{q}^{s}\right] \left( t/t_{0}\right) ^{-a}\right\} $ so
that the factor $\kappa \left( a\right) +K_{q}^{s}$ determines the sign and
the strength of the leading correction to the critical decay. For the HSS
one gets $\kappa \left( a\right) =-0.002$ \cite{Franosch97} and $%
K_{q}^{s}<K_{p}^{s}$ for $q<p $ (compare Fig. \ref{CritAmpR1}). This
explains in particular why at early times the curves with smaller wave
number fall below the curves with larger wave number. Since $\kappa \left(
a\right) +K_{q}^{s}$ changes sign between $q_{2}$ and $q_{3}$, the curves $0$%
, $1$, and $2$ fall below and the curve $3$ is above the dashed asymptote
for $t<10$. On the liquid side the analogous formula $\Phi _{q}^{s}\left(
t\right) =f_{q}^{sc}-h_{q}^{s}\tilde{t}^{b}\left\{ 1-\left[ \kappa \left(
-b\right) +K_{q}^{s}\right] \tilde{t}^{b}\right\} $ provides the explanation
for the deviations from the von Schweidler asymptote, Eq. (\ref{vSchwlaw}),
at later times: the $q$--dependence of the deviations is again determined by
the amplitude $K_{q}^{s}$. This entails that if one curve lies above the
other when deviating from the factorization theorem at short times it will
also do so at long times. This rule, which has already been discussed for $%
\Phi _{q}\left( t\right) $ in Ref. \cite{Franosch97}, holds for all
quantities that asymptotically obey Eq. (\ref{AE_SISF}). For the HSS one
gets $\kappa \left( -b\right) =0.569$ \cite{Franosch97} and therefore Fig. 
\ref{CritAmpR1} implies that $\kappa \left( -b\right) +K_{q}^{s}$ changes
sign between $q_{1} $ and $q_{2}$. This explains why the curves $0$ and $1$
are below, and $2$ and $3$ above the dashed asymptote in Fig. \ref{red_theo}
for $\log _{10}t\approx 6$. On the glass side the long--time behavior of $%
\Phi _{q}^{s}\left( t\right) $ is described by $%
f_{q}^{s}=f_{q}^{sc}+h_{q}^{s}\sqrt{\sigma /\left( 1-\lambda \right) }\left[
1+\sqrt{\sigma }\left( \bar{K}_{q}^{s}+\kappa \right) \right] $.

Armed with Fig. \ref{CritAmpR06} and what was explained for $d^{s}=1$, it is
easy to infer what Fig. \ref{red_theo} would look like for a hard--sphere of
diameter $d^{s}=0.6$. Since it was already explained in Ref. \cite
{Franosch97} how the deviations from the leading--order translate into the
frequency domain, spectra shall not be discussed in this paper.

In order to discuss the quality of the asymptotic expansions in a
controllable manner, some convention for the error margin has to be made.
There are various possibilities \cite{Franosch97}. In this paper the range
of validity shall be defined as the time window where for a given value $%
\Phi $ of the correlator, the time $t_{\Phi }$ for the solution, $\Phi =\Phi
(t_{\Phi })$, deviates from the time $t_{\Phi }^{as}$ for the asymptotic
expansion, say $\Phi =f_{q}^{c}+h_{q}G\left( t_{\Phi }^{as}\right) $, by
less than $\tilde{\varepsilon}$ per cent. This definition can always be
applied to monotonic functions of time, in particular to functions in the $%
\beta $--relaxation window. If one chooses $\tilde{\varepsilon}$ so small
that the deviations from the leading order are quantitatively explained by
the leading corrections, the range of validity for the leading--order $\beta 
$--relaxation asymptote can be shown to expand with the $\beta $ time scale $%
t_{\sigma }$ for $\sigma \rightarrow 0+$ and with the $\alpha $ time scale $%
t_{\sigma }^{\prime }$ for $\sigma \rightarrow 0-$. In Fig. \ref{red_theo}
and some of the following figures the endpoints of the time intervals with $%
\tilde{\varepsilon}=20\%$ are marked by various symbols.

\subsubsection{$\alpha $ relaxation\label{alphaphis}}

Within the $\alpha $--relaxation window the tagged--particle correlator $%
\Phi _{q}^{s}(t)$ obeys the superposition principle as formulated in Eq. (%
\ref{TDSP}) for the density correlator: $\Phi _{q}^{s}\left( t\right) =%
\tilde{\Phi}_{q}^{s}\left( t/t_{\sigma }^{\prime }\right) $. The
control--parameter--independent master function $\tilde{\Phi}_{q}^{s}\left( 
\tilde{t}\right) $ is obtained from Eq. (\ref{alpha_ISF}), where merely
superscripts $s$ have to be added to correlators and kernels \cite{Goetze91b}%
.

In Fig. \ref{ascale_phist} various rescaled correlators $\Phi _{1}^{s}(t)$
are compared to the $\alpha $ master function. The superposition principle
best describes the $\alpha $ relaxation for later times. As the glass
transition is approached from the liquid side, the range of validity of the
superposition principle, which is marked by diamonds and defined in analogy
to the $\beta $--relaxation regime above, extends to earlier rescaled times $%
\tilde{t}$. Close to the critical point, the deviations from the $\alpha $
master curve can be understood in terms of the leading--order correction $%
\delta \tilde{\Phi}_{q}^{s}\left( \tilde{t}\right) $ to the $\alpha $%
--scaling law, which is linear in $\sigma $: $\Phi _{q}^{s}(t)=\tilde{\Phi}%
_{q}^{s}\left( \tilde{t}\right) +\delta \tilde{\Phi}_{q}^{s}\left( \tilde{t}%
\right) +{\cal O}\left( \sigma ^{2}\right) $ \cite{Franosch97}. Although the
equation governing the time evolution of $\delta \tilde{\Phi}_{q}^{s}\left( 
\tilde{t}\right) $ is rather involved, its short--time behavior can be
expressed in terms of the amplitudes introduced above in connection with the 
$\beta $ relaxation. One gets up to errors of order $\sigma \tilde{t}^{b}$: 
\begin{equation}
\delta \tilde{\Phi}_{q}^{s}\left( \tilde{t}\right) =-\sigma h_{q}^{s}\left\{
B_{1}\tilde{t}^{-b}+\left[ \tilde{\kappa}\left( -b\right) -2B_{1}K_{q}^{s}-%
\hat{K}_{q}^{s}\right] \right\}   \label{alpha_corr}
\end{equation}
with $\tilde{\kappa}\left( -b\right) =2.97$ for the HSS \cite{Franosch97}.
The $\tilde{t}^{-b}$ term, which diverges for $\tilde{t}\rightarrow 0$,
dominates the deviations from $\alpha $ scaling for small $\left| \sigma
\right| $. Consequently, asymptotically close to $\varphi _{c}$, the
deviations occur at short times, where the $\alpha $ master function is
described by von Schweidler's law (\ref{vSchwlaw}). Therefore, the $\tilde{t}
$--window for the $\alpha $ scaling expands to smaller rescaled times
proportional to $\left| \sigma \right| ^{\frac{1}{2b}}$. For small $\left|
\sigma \right| $ the $\alpha $--relaxation window becomes independent of the
correction amplitudes $K_{q}^{s}$ and $\hat{K}_{q}^{s}$ and thus independent
of the observable under consideration, because the observable--dependent
amplitude $h_{q}^{s}$ occurs as a prefactor to both von Schweidler's law and
the relevant correction $\tilde{t}^{-b}$. The dotted line in Fig. \ref
{ascale_phist} illustrates how the deviations $\delta \tilde{\Phi}%
_{q}^{s}\left( \tilde{t}\right) $ from the $\alpha $ master function can be
described by the leading correction $-\sigma h_{q}^{s}B_{1}\tilde{t}^{-b}$.

\section{The mean--squared displacement\label{MSD}}

\subsection{Equation of motion}

The equation of motion for the mean--squared displacement $\delta
r^{2}\left( t\right) =\left\langle \left| \vec{r}\left( t\right) -\vec{r}%
\left( 0\right) \right| ^{2}\right\rangle $ \cite{Hansen86} can be obtained
from Eq. (\ref{MCE_SISF}) by exploiting its relation to the
small--wave--number behavior of the tagged--particle density correlator $%
\Phi _{q}^{s}\left( t\right) =1-q^{2}\delta r^{2}\left( t\right) /6+{\cal O}%
\left( q^{4}\right) $: 
\begin{equation}
\delta r^{2}\left( t\right) +D_{0}^{s}\int\limits_{0}^{t}m^{(0)}\left(
t-t^{\prime }\right) \delta r^{2}\left( t^{\prime }\right) dt^{\prime
}=6D_{0}^{s}t.  \label{MCE_MSD}
\end{equation}
Here we have introduced the kernel $m^{\left( 0\right) }\left( t\right)
=\lim_{q\rightarrow 0}q^{2}m_{q}^{s}\left( t\right) $. Carrying out the
limit $q\rightarrow 0$ in Eq. (\ref{MCF_SISF}) one finds the representation
of the new kernel as a new mode--coupling functional, $m^{\left( 0\right)
}\left( t\right) ={\cal F}_{MSD}\left( \Phi (t),\Phi ^{s}(t)\right) $, where 
\begin{equation}
{\cal F}_{MSD}\left( f,f^{s}\right) =\frac{1}{6\pi ^{2}}\int\limits_{0}^{%
\infty }nS_{k}c_{k}^{s2}k^{4}f_{k}f_{k}^{s}\,dk.  \label{MCF_MSD}
\end{equation}
Again this integral shall be rewritten as a Riemann sum over the previously
introduced wave--vector grid of $100$ terms 
\begin{equation}
{\cal F}_{MSD}\left( f,f^{s}\right) =n\Delta ^{5}/\left( 6\pi ^{2}\right)
\sum_{\hat{k}=\frac{1}{2}}^{99\frac{1}{2}}S_{k}c_{k}^{s2}\hat{k}%
^{4}f_{k}f_{k}^{s}.\,  \label{DMCF_MSD}
\end{equation}

Two side remarks concerning the preceding formulae might be of interest.
First, instead of solving Eq. (\ref{MCE_MSD}) with $\Phi _{q}(t)$ and $\Phi
_{q}^{s}(t)$ as an input for the determination of the kernel $m^{\left(
0\right) }$, one could obtain the desired result directly as the
small--wave--vector limit: $\delta r^{2}\left( t\right) =6\lim_{q\rightarrow
0}\left[ 1-\Phi _{q}^{s}\left( t\right) \right] /q^{2}$. We have studied
this procedure and checked the result to be the same as obtained from Eqs. (%
\ref{MCE_MSD}) and (\ref{MCF_MSD}). However, to do this, we had to use more
than $100$ grid points for the Riemann sum and had to extrapolate carefully
from small $q$ to zero. This is necessary, since the discretization in Eq. (%
\ref{DMCF_SISF}) is too crude to produce reliable results for very small $q$%
, say $q=\Delta $ or $2\Delta $. However, after carrying out the small--$q$
limit for $q^{2}m_{q}^{s}\left( t\right) $, the discretization (\ref
{DMCF_MSD}) of the integral in Eq. (\ref{MCF_MSD}) is harmless since the
major contributions come from the intermediate wave vector domain $k\sim 7$.
This reflects the fact that the sluggish dynamics of the tagged particle is
ruled by the cage effect, i.e., by structure correlations on length scales
of the interparticle distance.

Second, there is a trivial relation between $\delta r^{2}\left( t\right) $
and the velocity correlation function $K^{s}\left( t\right) $. Equations (%
\ref{MCE_MSD}) and (\ref{MCF_MSD}) are equivalent to the MCT equation
discussed earlier for $K^{s}\left( t\right) $ in the Laplace domain \cite
{Goetze91b}. Following this route $\delta r^{2}\left( t\right) $ has been
evaluated previously for some representative packing fractions for the HSS 
\cite{Fuchs95}. We prefer to solve Eq. (\ref{MCE_MSD}) directly, since
thereby we need not worry about a careful handling of the strong small--$s$
divergency exhibited by ${\cal L}\left[ \delta r^{2}\left( t\right) \right]
\left( s\right) $.

\subsection{The diffusion--localization transition}

In order to understand that the ideal liquid--glass transition implies a
transition from particle localization to particle diffusion, let us remember
that the Fourier back transform $\Phi ^{s}(r,t)=\left\langle \delta \left( 
\vec{r}-\left[ \vec{r}\left( t\right) -\vec{r}\left( 0\right) \right]
\right) \right\rangle $ of the tagged particle density correlator from the
wave--vector to the displacement domain is the probability density for
finding at time $t$ the particle in a distance $r$ from its starting
position. In the glass this distribution levels off for long times at the
normalized distribution $f^{s}\left( r\right) $. Hence the tagged particle
is localized in the glass matrix. A characteristic localization length $%
r_{s} $ can be defined, for example, in terms of the small--$q$ limit of the
Lamb--M\"{o}ssbauer factor: $f_{q}^{s}=1-\left( qr_{s}\right) ^{2}+{\cal O}%
\left( q^{4}\right) $. This fixes the long--time asymptote of the
mean--squared displacement for $\sigma >0$%
\begin{equation}
\lim_{t\rightarrow \infty }\delta r^{2}\left( t\right) =6r_{s}^{2}.
\end{equation}
Obviously, $\delta r^{2}\left( t\right) =\int \Phi ^{s}(r,t)r^{2}d^{3}r$, so
that $6r_{s}^{2}=\int f^{s}(r)r^{2}d^{3}r$. Equation (\ref{MCE_MSD}) yields
the general formula for $r_{s}$ in terms of sums over the glass form factors 
\cite{Goetze91b} 
\begin{equation}
r_{s}^{2}=1/{\cal F}_{MSD}\left( f,f^{s}\right) .  \label{critlocal}
\end{equation}
In the liquid state the low--frequency behavior of the Laplace transform of
all correlation functions and kernels is smooth \cite{Goetze95b}. Therefore,
one derives from Eq. (\ref{MCE_MSD}) the well known formula for the linear
divergency in time 
\begin{equation}
\lim_{t\rightarrow \infty }\delta r^{2}\left( t\right) /t=6D^{s}.
\end{equation}
Here $D^{s}$ is the tagged particle diffusivity. It is also called the
long--time diffusivity in order to distinguish it from $D_{0}^{s}$, which
determines the short--time asymptote $\lim_{t\rightarrow 0}\delta
r^{2}\left( t\right) /t=6D_{0}^{s}$. From Eq. (\ref{MCE_MSD}) one readily
derives for the ratio of $D^{s}$ and $D_{0}^{s}$%
\begin{equation}
\frac{D^{s}}{D_{0}^{s}}=\frac{1}{1+D_{0}^{s}\int_{0}^{\infty }m^{\left(
0\right) }\left( t\right) dt}.
\end{equation}
Note that this ratio is less than unity, which is an obvious manifestation
of the cage effect.

Figure \ref{r2} exhibits the evolution of the bifurcation dynamics as probed
by $\delta r^{2}\left( t\right) $ for the HSS. For very small $t$ one
observes short--time diffusion. With increasing density the short--time
diffusion is suppressed due to the cage effect. The glass curves level off
for long times at $6r_{s}^{2}$. With decreasing density the localization
length increases up to some critical value $r_{sc}$. The critical value $%
r_{sc}=0.0746$ fits nicely to the Lindemann melting criterion, as already
noticed in Ref. \cite{Bengtzelius84}. The liquid curves intersect the
plateau $r_{sc}$ for times of order $t_{\sigma }$; the time $t_{\sigma }$ is
indicated for the $n=9$ curve by a dot in Fig. \ref{r2}. Then they leave the
plateau according to von Schweidler's law in order to cross over to the
diffusion limit $\delta r^{2}\left( t\right) =6D^{s}t$ for times large
compared to $t_{\sigma }^{\prime }$. For the $n=9$ curve, the latter
asymptote is indicated by a dotted line in Fig. \ref{r2} and the time $%
t_{\sigma }^{\prime }$ is marked by a full square on the graph. The increase
of $\delta r^{2}\left( t\right) $ above $6r_{s}^{2}$ is the $\alpha $
process of the mean--squared--displacement dynamics. It deals with the
tagged particle's leaving of the cage. The initial part of this process,
where $\delta r^{2}\left( t\right) $ is close to $r_{sc}^{2}$, is stretched
over a large dynamical window if $\left| \varepsilon \right| $ is small. For
times of order $20t_{\sigma }^{\prime }$, $\delta r^{2}\left( t\right) $ has
increased to about unity; then stochastic dynamics sets in and $\delta
r^{2}\left( t\right) $ follows the diffusion asymptote. Between the end of
the regular short--time transient and the start of the $\alpha $ process,
there is a mesoscopic window for another anomalous dynamics. It deals with
the stretched approach of $\delta r^{2}\left( t\right) $ towards the plateau 
$6r_{s}^{2}$. In this sense the bifurcation dynamics, i.e., the dynamics
outside the transient, deals with a two--step relaxation process. Asymptotic
expansions shall be used to describe the bifurcation scenario by analytical
formulae, thereby providing an understanding of Fig. \ref{r2}.

\subsection{Asymptotic laws}

\subsubsection{$\beta $ relaxation}

An equation connecting the Laplace transforms $\delta r^{2}\left( s\right) $
and $m^{\left( 0\right) }\left( s\right) $ of $\delta r^{2}\left( t\right) $
and $m^{\left( 0\right) }\left( t\right) $, respectively, within the
structural relaxation regime $t\gg t_{0}$ follows from Eq. (\ref{SSR}): $%
s\delta r^{2}\left( s\right) =6/\left[ sm^{\left( 0\right) }\left( s\right)
\right] $. By inserting the asymptotic expansions (\ref{AE_ISF}) and (\ref
{AE_SISF}) into this equation one again obtains the general result for the $%
\beta $ relaxation to next--to--leading order 
\begin{equation}
\delta r^{2}\left( t\right) /6=r_{sc}^{2}-h_{MSD}G\left( t\right)
-h_{MSD}\left[ H\left( t\right) +K_{MSD}G\left( t\right) ^{2}+\sigma \hat{K}%
_{MSD}\right] .  \label{AE_MSD}
\end{equation}
Here $r_{sc}^{2}$ follows from Eq. (\ref{critlocal}) with $f$ and $f^{s}$
specialized to $f^{c}$ and $f^{sc}$, respectively, and the other amplitudes
read 
\begin{mathletters}
\label{AMP_MSD}
\begin{eqnarray}
h_{MSD} &=&r_{sc}^{4}\left[ {\cal F}_{MSD}^{c}\left( h,f^{sc}\right) +{\cal F%
}_{MSD}^{c}\left( f^{c},h^{s}\right) \right] , \\
K_{MSD} &=&\frac{r_{sc}^{4}}{h_{MSD}}\left[ {\cal F}_{MSD}^{c}\left(
h,h^{s}\right) +{\cal F}_{MSD}^{c}\left( hK,f^{sc}\right) +{\cal F}%
_{MSD}^{c}\left( f^{c},h^{s}K^{s}\right) \right] -\lambda \frac{h_{MSD}}{%
r_{sc}^{2}}, \\
\hat{K}_{MSD} &=&\frac{r_{sc}^{4}}{h_{MSD}}\left[ \frac{\partial {\cal F}%
_{MSD}^{c}\left( f^{c},f^{sc}\right) }{C\partial \varepsilon }+{\cal F}%
_{MSD}^{c}\left( h\hat{K},f^{sc}\right) +{\cal F}_{MSD}^{c}\left( f^{c},h^{s}%
\hat{K}^{s}\right) \right] -\frac{h_{MSD}}{r_{sc}^{2}},
\end{eqnarray}
where we have exploited that the mode--coupling functional ${\cal F}%
_{MSD}\left( f^{c},f^{sc}\right) $ is linear in $f$ and $f^{s}$ and
introduced the short--hand notation like $hK$ for $\left( hK\right)
_{p}=h_{p}K_{p}$.

In Fig. \ref{corr_r2} the mean--squared displacement is compared to the
asymptotic results in leading and in next--to--leading order for a tagged
particle in the HSS. The next--to--leading--order result explains the
deviations from the leading--order one. The general trend of the deviations
is again explained by the short--time corrections to the critical law and
the long--time corrections to the von Schweidler law on the liquid side---or
to the nonergodicity parameter $6r_{sc}^{2}$ on the glass side. The range of
validity is smaller than for any of the $q$--vectors discussed in connection
with Fig. \ref{red_theo}, since the absolute value of $K_{MSD}$ is larger
than of all the $K_{i}^{s}$ ($i=0,1,2,3$). The next--to--leading--order
results are a considerable improvement over the leading ones: On the glass
side, the second--order range of validity extends over almost three decades,
while the first--order range---as defined at the end of Sect. \ref{BETA_SISF}%
---does not even exist. On the liquid side the next--to--leading order adds
about a decade to the short--time side of the range of validity. It extends
it beyond the boundaries of the figure on the long--time side. However this
happens in part by accident because the exponent of the second term ($t^{2b}$%
, $2b=1.16$) in the von Schweidler series is close to the exact long--time
exponent, given by the diffusion law ($6D^{s}t$).

\subsubsection{$\alpha $ relaxation}

A description of the $\alpha $ process for the mean--squared--displacement
dynamics can be derived from Eq. (\ref{MCE_MSD}). Thereby one obtains the
superposition principle 
\end{mathletters}
\begin{equation}
\delta r^{2}\left( t\right) =\delta \tilde{r}^{2}\left( t/t_{\sigma
}^{\prime }\right) .  \label{AS_MSD}
\end{equation}
Here the master function $\delta \tilde{r}^{2}$ is to be calculated from 
\begin{equation}
\int\limits_{0}^{\tilde{t}}m^{(0)c}\left( \tilde{t}-t^{\prime }\right)
\delta \tilde{r}^{2}\left( t^{\prime }\right) dt^{\prime }=6\tilde{t},
\label{alpha_MSD}
\end{equation}
with the initial condition $\delta \tilde{r}^{2}\left( \tilde{t}\right)
=6\left[ r_{sc}^{2}+h_{MSD}\tilde{t}^{b}\right] +{\cal O}\left( \tilde{t}%
^{2b}\right) $. The kernel is given by the mode--coupling functional at the
critical point ${\cal F}_{MSD}^{c}$ and by the density--fluctuation master
functions: $m^{(0)c}\left( \tilde{t}\right) =$ ${\cal F}_{MSD}^{c}\left( 
\tilde{\Phi}\left( \tilde{t}\right) ,\tilde{\Phi}^{s}\left( \tilde{t}\right)
\right) $.

The various superposition principles imply coupling of the $\alpha $%
--relaxation time scales or relaxation rates in the following sense \cite
{Goetze91b}. Let us characterize the long--time decay of the variable $A$ in
the liquid by some time $\tau _{A}$. This time diverges upon approaching the
glass--transition: in the leading asymptotic limit for $\sigma \rightarrow
0- $ one finds $\tau _{A}=C_{A}t_{\sigma }^{\prime }$. All times or rates
are proportional to each other and follow a power law, specified by the
exponent $\gamma $: $1/\tau _{A}=\Gamma _{A}\left| \varepsilon \right|
^{\gamma }$, Eq. (\ref{timescales}). The constants of proportionality $C_{A}$
or $\Gamma _{A}$ depend on the variable $A$ and on the precise convention
for the definition of $\tau _{A}$. The scale coupling or $\alpha $--scale
universality is demonstrated in Fig. \ref{Diff_Coeff}, where the diffusion
coefficients of hard--sphere particles of diameters $d^{s}=1.0$ and $%
d^{s}=0.6$, and the $\alpha $--scaling rate $1/\tau _{1}^{\alpha s}$ of the
density fluctuations for wave vector $q_{1}$ are compared to the asymptotic
predictions. Although the asymptotic behavior is the same for all the
quantities, $D^{s}$ for $d^{s}=0.6$ already deviates visibly from the
asymptotic result for $n=4$, while the other quantities start to deviate
only for $n=3$. This underlines again the nonuniversality of the deviations.
Note that this difference might not be due to asymptotic corrections in the
sense discussed in this paper but could come from the mismatch of structural
relaxation, governed by the MCT kernel $m_{q}^{s}\left( t\right) $, and the
transient dynamics, ruled by the times $\tau _{q}^{s}$ in Eq. (\ref{MCE_SISF}%
).

In Fig. \ref{ascale_r2} the rescaled mean--squared displacement is compared
to the $\alpha $ master function $\delta \tilde{r}^{2}\left( \tilde{t}%
\right) $ for various packing fractions. The time scale $t_{\sigma }^{\prime
}$ is taken from Eq. (\ref{timescales}). Note that this figure is a harder
test of the asymptotics then just testing the time--density superposition
principle, since it tests the scaling time $t_{\sigma }^{\prime }$ in
addition to the shape of the master function.

The leading correction to (\ref{AS_MSD}) is known to be of order $\left|
\sigma \right| $ while the leading corrections to the factorization property
are of relative order $\sqrt{\left| \sigma \right| }$ \cite{Franosch97},
which explains why $\alpha $ scaling works much better than $\beta $
scaling. As we have already pointed out in Sect. \ref{alphaphis} the
''quality'' of $\alpha $ scaling is observable--independent---as can be seen
by comparing Fig. \ref{ascale_r2} to Fig. \ref{ascale_phist}---even though
first--order $\beta $ scaling is much worse for $\delta r^{2}\left( t\right) 
$ than for $\Phi _{1}^{s}\left( t\right) $. This is so because the dominant
correction to the long--time $\beta $ process is the corrections to the von
Schweidler law (see Fig. \ref{corr_r2}), which is absorbed into the
short--time expansion of the $\alpha $ master function. The dashed and
dotted curves in Fig. \ref{ascale_r2} show, how the deviations from the
superposition principle for small rescaled times $\tilde{t}$ are explained
by Eq. (\ref{alpha_corr}).

\section{Gaussian approximation\label{GA}}

The small--$q$ expansion of the density correlator $\Phi _{q}^{s}\left(
t\right) $, which lead us to the equation of motion for the mean--squared
displacement, can be viewed as the leading term of the Gaussian
approximation \cite{Hansen86,Rahman62,Rahman64} 
\begin{equation}
\Phi _{q}^{s}\left( t\right) \approx e^{-q^{2}\delta r^{2}\left( t\right)
/6}.  \label{EGA}
\end{equation}
The Gaussian approximation becomes exact for the short--time dynamics
(either ballistic or---as in our case---diffusive motion) and for the
long--time diffusion, and is known to work quite well for simple liquids
such as Argon in its normal state\cite{Rahman64}. So in the following we are
primarily interested in the intermediate time window of the
structural--relaxation regime.

In Fig. \ref{Gauss_Approx} some tagged--particle correlators as calculated
for $\left| \varepsilon \right| =10^{-7/3}$ are compared with the Gaussian
approximation. The latter describes the behavior of the self--intermediate
scattering function of the HSS reasonably for all considered wave numbers.
Coming from small wave numbers, at which the Gaussian approximation is
asymptotically exact, the deviations start to appear at the end of the $%
\beta $--relaxation regime, i.e., in the von Schweidler region as
exemplified for $q=q_{0}$. Going to still larger wave numbers like $q_{2}$
and $q_{3}$ the Gaussian approximation also deviates at the beginning of the 
$\beta $--relaxation regime, i.e., in the critical region.

The Gaussian approximation for the critical amplitudes can be found by
substituting Eq. (\ref{AE_MSD}) into Eq. (\ref{EGA}) and expanding the
exponential 
\begin{mathletters}
\label{GA_Crit_Amp}
\begin{eqnarray}
f_{q}^{sc} &\approx &e^{-q^{2}r_{sc}^{2}},  \label{1GA} \\
h_{q}^{s} &\approx &e^{-q^{2}r_{sc}^{2}}q^{2}h_{MSD},  \label{2GA} \\
K_{q}^{s} &\approx &K_{MSD}+\frac{1}{2}q^{2}h_{MSD},  \label{3GA} \\
\bar{K}_{q}^{s} &\approx &\bar{K}_{MSD}+\frac{1}{2\sqrt{1-\lambda }}%
q^{2}h_{MSD}.  \label{4GA}
\end{eqnarray}
Here $\bar{K}_{MSD}$ has been introduced in analogy to Eq. (\ref{Kbar}).
These approximation results are compared to the various amplitudes in Fig. 
\ref{CritAmpR1} and \ref{CritAmpR06}. In both cases we find a good
qualitative description for the amplitudes; the Gaussian approximation even
describes the change of sign of $K_{q}^{s}$ and $\bar{K}_{q}^{s}$. The
Gaussian results for $f^{sc}$ and $h^{s}$ are in better quantitative
agreement with the exact results than the results for $K_{q}^{s}$ and $\bar{K%
}_{q}^{s}$, since the leading small--$q$ corrections to (\ref{1GA}) and (\ref
{2GA}) are of higher order than those for (\ref{3GA}) and (\ref{4GA}).
Obviously the results for $d^{s}=1.0$ are better than for the smaller
particle with $d^{s}=0.6$. For the large particle they even give good
quantitative descriptions for some of the amplitudes: practically no
deviation can be found for the critical Lamb--M\"{o}ssbauer factor $f^{sc}$
up to $q=15$, and the critical amplitude $h^{s}$ is reproduced up to $q=5.$

These results also explain the deviations for the dynamics in Fig. \ref
{Gauss_Approx}: Since the critical amplitude $h^{s}$ is overestimated by the
Gaussian approximation, the dashed curves relax faster in the $\beta $%
--relaxation region. On the glass side, a larger $h^{s}$ leads to a larger $%
f^{s}$ via $f^{s}=f^{sc}+h^{s}\sqrt{\sigma /\left( 1-\lambda \right) }$. In
the same way the corresponding figure for $d^{s}=0.6$ can be inferred from
Fig \ref{CritAmpR06}.

\section{non--Gaussian parameter\label{NGP}}

\subsection{The equation of motion}

A cumulant expansion can be used to systematically classify the deviations
of the density correlator from its Gaussian approximation. The leading
contribution to the deviations is proportional to the non--Gaussian
parameter $\alpha _{2}\left( t\right) $. One gets 
\end{mathletters}
\begin{equation}
\Phi _{q}^{s}\left( t\right) =e^{-q^{2}\delta r^{2}\left( t\right)
/6}\left\{ 1+\frac{1}{2}\alpha _{2}\left( t\right) \left[ q^{2}\delta
r^{2}\left( t\right) /6\right] ^{2}+{\cal O}\left( q^{6}\right) \right\} ,
\label{cumulant}
\end{equation}
where $\alpha _{2}=\left( 3/5\right) \left[ \delta r^{4}\left( t\right)
/\delta r^{2}\left( t\right) ^{2}\right] -1$ with $\delta r^{4}\left(
t\right) =\left\langle \left| \vec{r}\left( t\right) -\vec{r}\left( 0\right)
\right| ^{4}\right\rangle $ \cite{Hansen86,Rahman62,Rahman64}. Since the
mean--quartic displacement $\delta r^{4}\left( t\right) $ is proportional to
the fourth Taylor coefficient in a small--$q$ expansion of $\Phi
_{q}^{s}\left( t\right) =1-q^{2}\delta r^{2}\left( t\right) /3!+q^{4}\delta
r^{4}\left( t\right) /5!+\ldots $, one can derive an equation of motion for $%
\alpha _{2}$ from Eqs. (\ref{MCE_SISF}) and (\ref{MCE_MSD}) 
\begin{equation}
\left[ 1+\alpha _{2}\left( t\right) \right] \delta r^{2}\left( t\right)
^{2}+D_{0}^{s}\int\limits_{0}^{t}m^{\left( 0\right) }\left( t-t^{\prime
}\right) \left[ 1+\alpha _{2}\left( t^{\prime }\right) \right] \delta
r^{2}\left( t^{\prime }\right) ^{2}dt^{\prime
}=6D_{0}^{s}\int\limits_{0}^{t}\left[ 2+m^{\left( 2\right) }\left(
t-t^{\prime }\right) \right] \delta r^{2}\left( t^{\prime }\right)
dt^{\prime }.  \label{MCE_NGP}
\end{equation}
Here a further mode--coupling functional for the kernel $m^{\left( 2\right)
}\left( t\right) ={\cal F}_{NGP}\left( \Phi \left( t\right) ,\Phi ^{s}\left(
t\right) \right) $ is introduced: 
\begin{equation}
{\cal F}_{NGP}\left( f,f^{s}\right) =\frac{1}{10\pi ^{2}}%
\displaystyle \int %
nS_{k}c_{k}^{s2}k^{4}f_{k}\left( \frac{\partial ^{2}f_{k}^{s}}{\partial k^{2}%
}+\frac{2}{3k}\frac{\partial f_{k}^{s}}{\partial k}\right) dk.
\label{MCF_NGP}
\end{equation}
For a numerical calculation the integral (\ref{MCF_NGP}) is expressed by a
Riemann sum analogous to Eq. (\ref{DMCF_MSD}). The derivatives $\partial
f_{k}^{s}/\partial k$ and $\partial ^{2}f_{k}^{s}/\partial k^{2}$ are
approximated through their numerical equivalents, e.g. $\partial
f_{k}^{s}/\partial k=\left( f_{k+\Delta }-f_{k-\Delta }\right) /\left(
2\Delta \right) $, except for $q=0.2$ and $q=0.6$, where they are calculated
from a small--$q$ extrapolation according to the Gaussian approximation. For
the same reasons as discussed above in connection with Eqs. (\ref{MCE_MSD}),
(\ref{MCF_MSD}), and (\ref{DMCF_MSD}) it is preferable to solve the derived
equation for $\alpha _{2}\left( t\right) $ directly rather than to deduce
the non--Gaussian parameter as the small--$q$ limit from the numerical
results for $\Phi _{q}^{s}\left( t\right) $. We have nevertheless checked
that the latter procedure can be followed provided one chooses a finer wave
vector grid and a proper large--$q$ regularization of the integral (\ref
{MCF_SISF}).

\subsection{The transition scenario}

To begin with, some comments concerning the general properties of $\alpha
_{2}\left( t\right) $ might be appropriate. Neither $\delta r^{2}\left(
t\right) $ nor $\alpha _{2}\left( t\right) $ are correlation functions and,
therefore, the known general properties of positive definite functions like $%
\Phi _{q}\left( t\right) $ and $\Phi _{q}^{s}\left( t\right) $ need not be
valid for these quantities. With $A\left( t\right) =\left| \vec{r}\left(
t\right) -\vec{r}\left( 0\right) \right| ^{2}$ denoting a positive
observable, one gets $\delta r^{2}\left( t\right) =\left\langle A\left(
t\right) \right\rangle $ and therefore $\delta r^{2}\left( t\right) \geq 0$.
Since $\Phi _{q}^{s}\left( t\right) =1-q^{2}\delta r^{2}\left( t\right) /6+%
{\cal O}\left( q^{4}\right) $ one finds for completely monotone density
correlators: $-\left( -\partial /\partial t\right) ^{l}\delta r^{2}\left(
t\right) \geqslant 0$ for $l=1,2,\ldots $ So within the structural
relaxation window $t\gg t_{0}$, and for all $t$ for our colloid model
defined by Eqs. (\ref{MCE_ISF}) and (\ref{MCE_SISF}), the mean--squared
displacement is an increasing function of time: $\left( \partial /\partial
t\right) \delta r^{2}\left( t\right) \geq 0$. The inequality $\left\langle
A^{2}\right\rangle \geq \left\langle A\right\rangle ^{2}$ implies $\alpha
_{2}\left( t\right) \geq -2.5$. Hence, the non--Gaussian parameter is bounded
from below, but it can have either sign. At early times the density
correlator exhibits Gaussian behavior, independent of whether Newtonian or
Brownian dynamics is considered, and therefore $\alpha _{2}\left(
t\rightarrow 0\right) =0$. The long--time liquid correlators describe
diffusion, which is a Gaussian process, and thus $\alpha _{2}\left(
t\rightarrow \infty \right) =0$. From Eqs. (\ref{MCE_NGP}) and (\ref{MCE_MSD}%
) one finds for $\sigma <0$ that $\alpha _{2}\left( t\rightarrow \infty
\right) ={\cal O}\left( 1/t\right) $. Hence, in the liquid state $\alpha
_{2}\left( t\right) $ cannot be a monotonic function of time. Negative $%
\alpha _{2}\left( t\right) $ means, that the probability for the particle to
move very far is suppressed relative to the one expected for a random--walk
process. Similarly, the probability for moving far is enhanced if $\alpha
_{2}\left( t\right) >0$. In the latter case the cage boundary is more fuzzy
than in the former.

Figure \ref{alpha2} exhibits the evolution of the non--Gaussian parameter $%
\alpha _{2}\left( t\right) $ upon crossing the bifurcation point for tagged
spheres of diameters $d^{s}=1.0$ and $d^{s}=0.6$ in the HSS. For the glass
state the curves exhibit arrest for late times $\alpha _{2}\left(
t\rightarrow \infty \right) =f_{NGP}=1+{\cal F}_{NGP}\left( f,f^{s}\right) $%
. For the liquid state the $\alpha $ process manifests itself as a bump of
the $\alpha _{2}$--versus--$\log _{10}t$ diagram starting with an increase
above the plateau $f_{NGP}^{c}$, reaching some maximum, and then decreasing
to zero. The liquid curves for $n\geq 10$, i.e. for $\left| \varepsilon
\right| <10^{-3}$, exhibit the superposition principle: a change of $\sigma $
causes a shift of the peak parallel to the logarithmic time axis without
change of the shape. The first structural relaxation step deals with the
approach towards $f_{NGP}^{c}$ from below for $t\gg t_{0}$. The functions
for $\left| \varepsilon \right| \leq 10^{-4}$ exhibit the
two--step--relaxation scenario for $t\gtrsim 10^{4}$.

For dense normal liquids $\alpha _{2}$ is found to be positive in agreement
with molecular dynamics results for liquid argon near its triple point. Our
results for $d^{s}=1.0$ and $n=2$ are of about the same magnitude as the
value $0.13$ found for this Lennard--Jones system \cite{Rahman64}.
Increasing the density towards the critical one implies an increase of $%
\alpha _{2}$ by about a factor three for both values of $d^{s}$ studied; but
for the smaller spheres the non--Gaussian parameter is about three times
larger than for the larger ones. One concludes that glassy dynamics in the
liquid does not lead to dramatic changes of the magnitude of the
non--Gaussian parameter, a finding also supported by Fig. \ref{Gauss_Approx}.

For long times, and for the smaller sphere for all times, $\alpha _{2}$ is
positive. This enhanced probability for the particle to move further is what
one would expect as a result of the building of a backflow pattern in the
liquid. However, for $d^{s}=1.0$ the negative plateau value $f_{NGP}^{c}$
implies, that for sufficiently large densities a dip to negative values for $%
\alpha _{2}\left( t\right) $ appears. For $d^{s}=0.6$ the plateau is
positive and the dip does not exist. The depth of the predicted dip, as
opposed to the height of the maximum in the $\alpha $ regime, is not a
structural relaxation phenomenon. Rather it is caused by the cross over from
the short--time transient to the first structural relaxation step. The
mode--coupling approximations are constructed to describe the long--time
behavior due to the cage effect. The theory does not handle short--time
collision effects correctly, and therefore the size of the dip predicted for 
$d^{s}=1.0$ might be an artifact due to the insufficiencies of the MCT in a
regime, which it has not been made for. An analogous reservation applies, of
course, to the quantitative details of the crossover regimes displayed in
Fig. \ref{phis of t} for $\Phi _{q}^{s}\left( t\right) $ and in Fig. \ref{r2}
for $\delta r^{2}\left( t\right) $. The MCT equations do not guarantee the
validity of the inequality $\alpha _{2}\left( t\right) \geq -2.5$. Indeed
for $d^{s}=1.0$ and $\varphi >\varphi _{c}+0.01$ the inequality is violated
and therefore the corresponding glass curves are not shown in Fig. \ref
{alpha2}.

The extension of the Gaussian approximation according to Eq. (\ref{cumulant}%
) is illustrated by the dashed dotted lines in Fig. \ref{Gauss_Approx}. For
the wave vectors $q\leq q_{2}$, where the Gaussian approximation yields a
good description of the correlators, the addition of the leading--cumulant
correction proportional to $\alpha _{2}\left( t\right) $ improves the fit
seriously. However, for larger wave vectors the addition of the $\alpha _{2}$%
--term does not lead to improvements as is demonstrated for wave vector $%
q_{3}$ in Fig. \ref{Gauss_Approx}. As $\alpha _{2}$ is considerably larger
for $d^{s}=0.6$ than for $d^{s}=1.0$ the cumulant expansion (\ref{cumulant})
already breaks down for $q$ near $q_{0}$.

Figures \ref{phis of t}, \ref{r2}, and \ref{alpha2} have been shown in order
to illustrate the theoretical essence of the MCT bifurcation. To avoid
misleading conclusions from Fig. \ref{alpha2} it might be adequate to
remember the windows accessible by state--of--the--art experimental studies.
In molecular--dynamics work \cite{Kob95,Sciortino96} a variation of the
diffusivity or other structural relaxation scales could be detected over a
window of about four decades. This corresponds to the curves $n\lesssim 7$
in our figures. The dynamical window for structural relaxation explored by
van Megen and Underwood \cite{Megen94b} by photon--correlation spectroscopy
is also about four decades wide. The size of the dynamical window accessible
by the neutron--spin--echo instrument \cite{Mezei91} is smaller than three
decades. To focus on results for $\alpha _{2}$ which might be relevant for
the interpretation of experiments with the techniques available today, one
should ignore the results for $n\gtrsim 8$ in Fig. \ref{alpha2}. This
restricted set of results does neither exhibit the superposition principle
nor the two--step relaxation scenario. The leading--order asymptotic results
for the MCT bifurcation dynamics do not describe the MCT results shown in
Fig. \ref{alpha2} for $n\lesssim 8$, i.e., for $\left| \varepsilon \right|
>0.001$, not even qualitatively. But in the following it shall be shown that
the results can be understood in terms of the next--to--leading asymptotic
laws.

\subsection{Asymptotic laws}

\subsubsection{$\beta $ relaxation}

In order to work out the dynamics in the $\beta $--relaxation window we
start from the Laplace transform of Eq. (\ref{MCE_NGP}) 
\begin{equation}
{\cal L}\left[ \left( 1+\alpha _{2}\left( t\right) \right) \delta
r^{2}\left( t\right) ^{2}\right] \left( s\right) =s\delta r^{2}\left(
s\right) ^{2}\left\{ 2+s{\cal L}\left[ {\cal F}_{NGP}\left( \Phi \left(
t\right) ,\Phi ^{s}\left( t\right) \right) \right] \left( s\right) \right\} ,
\end{equation}
where the Laplace transform of Eq. (\ref{MCE_MSD}) has been used. Inserting
the asymptotic expansions (\ref{AE_ISF}), (\ref{AE_SISF}), and (\ref{AE_MSD}%
) for $\Phi _{q}\left( t\right) $, $\Phi _{q}^{s}\left( t\right) $, and $%
\delta r^{2}\left( t\right) $, respectively, one gets: 
\begin{equation}
\alpha _{2}\left( t\right) =f_{NGP}+h_{NGP}G\left( t\right) +h_{NGP}\left[
H\left( t\right) +K_{NGP}G\left( t\right) ^{2}+\sigma \hat{K}_{NGP}\right] .
\label{AE_NGP}
\end{equation}
Here the amplitudes are given by 
\begin{mathletters}
\label{AMP_A2}
\begin{eqnarray}
h_{NGP} &=&{\cal F}_{NGP}^{c}\left( h,f^{sc}\right) +{\cal F}%
_{NGP}^{c}\left( f^{c},h^{s}\right) , \\
K_{NGP} &=&\left[ {\cal F}_{NGP}^{c}\left( h,h^{s}\right) +{\cal F}%
_{NGP}^{c}\left( hK,f^{sc}\right) +{\cal F}_{NGP}^{c}\left(
f^{c},h^{s}K^{s}\right) \right] /h_{NGP}  \nonumber \\
&&+\frac{h_{MSD}}{r_{sc}^{4}}\left( 1-\lambda \right) \left[
2r_{sc}^{2}h_{NGP}-h_{MSD}(1+f_{NGP})\right] /h_{NGP}, \\
\hat{K}_{NGP} &=&\left[ \frac{\partial {\cal F}_{NGP}^{c}\left(
f^{c},f^{sc}\right) }{C\partial \varepsilon }+{\cal F}_{NGP}^{c}\left( h\hat{%
K},f^{sc}\right) +{\cal F}_{NGP}^{c}\left( f^{c},h^{s}\hat{K}^{s}\right)
\right] /h_{NGP}  \nonumber \\
&&+\frac{h_{MSD}}{r_{sc}^{4}}\left[
h_{MSD}(1+f_{NGP})-2r_{sc}^{2}h_{NGP}\right] /h_{NGP}.
\end{eqnarray}

The analytic results are compared with the numerical solutions for $n=9$ in
Fig. \ref{corr_a2}. Because of the large negative value of $h_{NGP}$, the
leading order qualitatively accounts for the steep rise of $\alpha _{2}$ in
the $\beta $--relaxation region. This is a major reason for the absence of a
plateau even though the reduced packing fraction $\left| \varepsilon \right|
=0.001$ is so small. The large size of $h_{NGP}$ also explains, why the
long--time limit $f_{NGP}$ in the glass is much more suppressed below the
plateau value $f_{NGP}$, than shown in Fig. \ref{corr_r2} for $\delta
r^{2}\left( t\right) $. There are two further peculiarities hidden in Eqs. (%
\ref{AMP_A2}), which render the discussion of Fig. \ref{corr_a2} different
from those considered so far for other examples. The $\alpha $ process for
the density correlators and for the mean--squared displacement deals with
monotonic functions. In those cases the leading corrections to the von
Schweidler law influence the details but not the general trend of the
functions. The positive slope of $\alpha _{2}\left( t\right) $ in the
leading--order--$\beta $--relaxation regime, however, has to change to a
negative one in the late $\alpha $--relaxation regime since the $\alpha
_{2}\left( t\right) $--versus--$t$ curve eventually has to decrease to zero.
Therefore the $\alpha _{2}\left( t\right) $--versus--$\log _{10}t$ diagram
exhibits a bump. The next--to--leading asymptotic formula (\ref{AE_NGP}), as
opposed to the leading one, can reproduce at least qualitatively this new
feature of the diagram as shown by the dot--dashed lines in Fig \ref{corr_a2}%
. The second difference concerns the remarkable parallel shift of the
solution relative to the leading asymptote occurring for times near $%
t_{\sigma }^{-}$ where the liquid $\beta $ correlator exhibits a zero, $%
G\left( t_{\sigma }^{-}\right) =0$ ($t_{\sigma }^{-}=0.704t_{\sigma }$ for
the HSS). The correction is given by $h_{NGP}\left[ H\left( t_{\sigma
}^{-}\right) +\sigma \hat{K}_{NGP}\right] $. For the previously discussed
examples the two terms in the bracket partly cancelled each other while for $%
\alpha _{2}\left( t\right) $ they add up. As a result this shift is about an
order of magnitude larger than in the examples studied in Figs. \ref
{red_theo} and \ref{corr_r2}, or in the examples considered in Ref. \cite
{Franosch97}. This shift is responsible for the fact that for $n=9$ the
leading--order asymptotic result for $\alpha _{2}\left( t\right) $ ($d^{s}=1$%
) (dashed lines in the lower panel of Fig. \ref{corr_a2}) has a smaller
range of validity than the corresponding result for the mean--square
displacement (Fig. \ref{corr_r2}) even though $|K_{NGP}|<|K_{MSD}|$. Of course,
this does not contradict the idea that the amplitude $K$ asymptotically
determines the range of validity of the leading--order result, because,
asymptotically close to the critical point, the shift becomes irrelevant.
Thus, the leading--order range of validity for $\alpha _{2}\left( t\right) $
($d^{s}=1$) will finally overtake that for $\delta r^{2}\left( t\right) $ as 
$\sigma \rightarrow 0$.

One can substitute the expansions (\ref{AE_MSD}) and (\ref{AE_NGP}) into Eq.
(\ref{cumulant}) in order to obtain the amplitudes in Eq. (\ref{AE_SISF}).
The result extends the Gaussian approximation (\ref{GA_Crit_Amp}) for these
quantities so that the first cumulant is taken care of. The result improves
the Gaussian approximation for $q\lesssim q_{0}$. However, for large wave
vectors the extended approximation is worse than the Gaussian one, which is
shown in Fig. \ref{CritAmpR1} and \ref{CritAmpR06} by the full lines.
Therefore the described extension is not worthwhile.

\subsubsection{$\alpha $ relaxation}

To explore the asymptotic behavior in the $\alpha $--relaxation region, the
non--Gaussian parameter is plotted versus rescaled time $\tilde{t}%
=t/t_{\sigma }^{\prime }$ in Fig. \ref{ascale_a2}. Comparing the results for 
$d^{s}=1.0$ with the corresponding ones for the mean--squared displacement
in Fig. \ref{ascale_r2} one notices, that one has to choose a considerably
smaller separation $\left| \sigma \right| $ for $\alpha _{2}\left( t\right) $
than for $\delta r^{2}\left( t\right) $ to find comparable agreement with
the $\alpha $--relaxation scaling law. For $n=5$ and $\tilde{t}=1$, $\delta
r^{2}\left( t\right) $ agrees with the superposition--principle asymptote
within the accuracy of the drawing in Fig. \ref{ascale_r2}, while the
corresponding result for $\alpha _{2}$ deviates seriously from the
asymptote. The mean--squared displacement is dominated by the diffusion
limit, i.e. Gaussian behavior for $\tilde{t}>20$. Such contribution is
absent in $\alpha _{2}$ and therefore the $n=5$ curve in Fig. \ref{ascale_a2}
magnifies the small deviations from Gaussian dynamics, which for the $n=5$
curve in Fig. \ref{ascale_r2} start to be visible only for $\tilde{t}%
\lesssim 0.3$.

Figures \ref{ascale_a2} and \ref{ascale_r2} are seemingly in contradiction
to the statement that $\alpha $--scaling should work equally well for
different quantities (see Sect. \ref{alphaphis}). However, as mentioned
above, the non--Gaussian parameter for the shown packing fractions is
special in the sense, that the corrections for small $\tilde{t}$ contain an
important shift term, which stems from the term in brackets in the
short--time expansion of the $\alpha $--scaling correction (\ref{alpha_corr}%
). In Fig. \ref{ascale_a2} this finding is illustrated for the curves with
label $n=9$. For even smaller $\left| \sigma \right| $ this shift becomes
irrelevant, thus resolving the paradox.

\section{Conclusions}

In this paper the MCT results for the structural relaxation of the
mean--squared displacement $\delta r^{2}\left( t\right) $ and non--Gaussian
parameter $\alpha _{2}\left( t\right) $ were discussed for the hard--sphere
system. More generally, we studied the generating function of these
quantities, viz the incoherent intermediate scattering function or
tagged--particle correlator $\Phi _{q}^{s}\left( t\right) $. The work is
motivated by the distinguished role of these quantities for the
interpretation of spectroscopic data and molecular--dynamics simulations of
simple glass forming liquids. The essence of the MCT bifurcation scenario
for the evolution of structural relaxation is contained in the leading
asymptotic laws which deal with the two scaling laws reviewed in Sect. \ref
{GTHSM}. Most of the so far published tests of MCT focused on an assessment
of these universal results. The outcome of this paper demonstrates in a
drastic manner, that the range of validity of the mentioned asymptotic laws
is not universal, rather it depends on the quantity considered. This range
of validity can be determined by working out the laws for the leading
corrections. Thereby formulae for a refined data analysis are obtained.
There are quantities such as $\alpha _{2}\left( t\right) $, which are not
described at all within the presently accessible dynamical window by the
leading asymptotic result. In this case the laws including the
leading--order corrections are necessary for a qualitative understanding.

From Fig. \ref{red_theo} one infers that it is relatively easy to extract
von Schweidler's law for the relaxation at the structure--factor peak
position $q_{1}$, while it is more difficult to identify this leading
asymptotic law, Eq. (\ref{vSchwlaw}), for the larger wave vector $q_{3}$.
The behavior of $\Phi _{q}^{s}\left( t\right) $ for the intermediate
wave--vector range $q_{0}$--$q_{3}$ is quite similar to what was studied for
the coherent intermediate scattering function in Ref. \cite{Franosch97}. The
results, in particular those for the spectra, can therefore be inferred from
that earlier work, using the amplitudes from Figs. \ref{CritAmpR1} and \ref
{CritAmpR06}. Inclusion of the calculated correction terms often extends the
range of validity of the asymptotic formulae seriously. Therefore it is
advisable to fit the initial part of the $\alpha $ process by the von
Schweidler expansion including the $\tilde{t}^{2b}$ term as was done
recently by Sciortino et al. \cite{Sciortino96} in their analysis of
simulation data for supercooled water. These authors noticed in particular,
that the leading correction term to Eq. (\ref{vSchwlaw}) was smallest near $%
q_{1}$. This is in qualitative agreement with Fig. \ref{red_theo}, even
though there is no a priori reason to expect our hard--sphere--system
results to be relevant for an explanation of the details of water.

Fig. \ref{corr_r2} implies that the $\delta r^{2}\left( t\right) $--versus--$%
\log _{10}t$ diagram exhibits only a small window for von Schweidler's law
and that the window for the analytic description is extended significantly
by inclusion of the $\tilde{t}^{2b}$ correction term. It is also predicted
that the critical decay law cannot be identified for reduced packing
fractions $\left| \varepsilon \right| \geq 0.001$. These findings are
similar to what was reported in Ref. \cite{Kob95} for the simulation results
for a binary Lennard--Jones mixture.

In agreement with photon--correlation--spectroscopy results for a
hard--sphere colloid \cite{Megen98} one infers from Fig. \ref{Gauss_Approx}
and from the lower panel of Fig. \ref{alpha2}, that the Gaussian
approximation works very well for a hard--sphere system. This finding is not
in conflict with the observation that $\delta r^{2}\left( t\right) $ does
not exhibit the critical decay law, while $\Phi _{q}^{s}\left( t\right)
=\exp \left[ -q^{2}\delta r^{2}\left( t\right) /6\right] $ does show this
leading asymptotic result, Eq. (\ref{critlaw}), for $q\sim q_{3}$ (see Fig. 
\ref{red_theo}). The strong $t^{-2a}$ corrections to $\delta r^{2}\left(
t\right) /6\sim r_{sc}^{2}-h_{MSD}\left( t_{0}/t\right) ^{a}$ nearly cancel
those coming from the expansion of $\exp \left[ -q^{2}\delta r^{2}\left(
t\right) /6\right] $; the relevant correction amplitude $K_{MSD}$ for $%
\delta r^{2}\left( t\right) $ is therefore larger than that for $\Phi
_{3}^{s}\left( t\right) $.

The test of MCT against molecular--dynamics--simulation results by Kob and
Andersen \cite{Kob95,Kob95b,Kob95c} gives strong support of the theory. This
can be appreciated even more if one considers the comparisons of
first--principle MCT calculations of the critical form factors, critical
amplitudes, and exponent parameter with the data \cite{Nauroth97}. The
authors also confirmed the power law singularity for the $\alpha $%
--relaxation scales $\tau \propto \left| T-T_{c}\right| ^{-\gamma }$ with
the predicted connection between the exponent $\gamma $ and the von
Schweidler exponent $b$, Eq. (\ref{timescales}). However, they obtained the
result for the diffusivity $D\propto \left| T-T_{c}\right| ^{\gamma ^{\prime
}}$ with $\gamma ^{\prime }<\gamma $, in contradiction to the $\alpha $%
--scale universality. This observation underlines in particular that the
predicted $\alpha $--scale coupling is not a triviality. Figure \ref
{Diff_Coeff} demonstrates that the corrections to the leading asymptotic law
for the diffusivity of a smaller sphere are larger than the ones for the $%
\alpha $--relaxation scale for a representative intermediate wave vector $%
q_{1}$. But our finding for the corrections to the hard--sphere--system
asymptotics are too small to explain the reported results for the cited
mixture. The MCT prediction for the universal $\alpha $--scales for the HSS, 
$\tau _{\alpha }\propto \left[ \left| \varphi -\varphi _{c}\right| /\varphi
_{c}\right] ^{-2.6}\propto 1/D$, was confirmed by a recent analysis of the
diffusivity $D$ and of the tagged--particle--correlator $\alpha $ scale $%
\tau _{\alpha }$ for a hard--sphere colloid by van Megen et al. \cite
{Megen98}.

The results for $\alpha _{2}\left( t\right) $ in the upper panel of Fig. \ref
{alpha2} are unusual if compared to the corresponding ones for
representative density correlators $\Phi _{q}\left( t\right) $ or $\Phi
_{q}^{s}\left( t\right) $. For the curves with $n\lesssim 8$, dealing with
the presently in experiments or simulations accessible parameters and
windows, the two MCT scaling laws, Eqs. (\ref{beta_scaling}) and (\ref{TDSP}%
), cannot be identified. Neither is it possible to recognize the two time
fractals, Eqs. (\ref{critlaw}) and (\ref{vSchwlaw}). These findings can be
understood by analytical formulae only, if the next--to--leading asymptotic
results are appreciated, as is shown in Fig. \ref{corr_a2}. However, these
observations are not so surprising for the following reasons. The $\alpha $
process for the correlator $\Phi _{q}\left( t\right) $ deals with the
monotonic decay from the plateau $f_{q}^{c}$ to zero. This phenomenon is
described already within Maxwell's visco--elastic theory, dealing with
exponential decay and Gaussian density fluctuations. Glassy dynamics deals
with deviations from this phenomenological description. However, for the
time range where $\Phi _{q}^{s}\left( t\right) <f_{q}^{sc}/2$, which deals
with a major part of the $\Phi _{q}^{s}\left( t\right) $--versus--$\log
_{10}t$ graph, the phenomenological picture remains essentially valid. In
this part the superposition principle holds already for $n\geq 5$, as can be
inferred from the $\alpha $--scaling analysis in Fig. \ref{ascale_phist} or
from Ref. \cite{Franosch97}. Also for $\alpha _{2}\left( t\right) $ the
superposition principle works for the mentioned large times as shown in Fig. 
\ref{ascale_a2}. But since there Gaussian dynamics is nearly valid, $\alpha
_{2}\left( t\right) $ is small and the corresponding part of the $\alpha
_{2} $--versus--$\log _{10}t$ diagram merely deals with a not so interesting
feature of the figure. The diagram is dominated by the maximum and this is
located near the end of the window of von Schweidler's law. Here the
corrections are large compared to the leading terms, the maximum is only
poorly approximated for $n\lesssim 5$. But Fig. \ref{ascale_phist} shows
that for $\Phi _{q}^{s}\left( t\right) $ the $n=5$ results do not follow the 
$\alpha $--scaling master curve for $t/t_{\sigma }^{\prime }<0.2$. This too
can be inferred from the corresponding scaling plot shown as Fig. $17$ in
Ref. \cite{Franosch97} or from Fig. \ref{ascale_r2} for $\delta r^{2}\left(
t\right) $. In the latter case the deviations from scaling are judged
relative to the underlying elementary background $f_{q}^{c}$ and $%
6r_{sc}^{2} $, respectively, and therefore they do not appear as qualitative
effects. The non--Gaussian parameter magnifies a small effect, and therefore
its approximation by asymptotic formulae is judged differently than for the
other functions. Indeed, for the HSS the predicted size $\alpha _{2}\left(
t\right) <0.3$ is so small, that the effect disappears in the data noise of
the measurement on colloids \cite{Megen98}. Let us emphasize that the value $%
f_{NGP}$ of the plateau was of importance for the preceding discussion.
Therefore, it would be of great interest for an assessment of our results to
measure $f_{NGP}$ in the glass state.

The upper panel of Fig. \ref{alpha2} looks similar to what Kob and Andersen
reported for their results on a binary mixture \cite{Kob95}. In particular
we also find $\alpha _{2}$ to be larger for smaller particles; the maximum
increases from $1.0$ to $1.2$ if $d^{s}$ is decreased from $0.6$ to $0.5$.
The clustering of the $\alpha _{2}$--versus--$\log _{10}t$ graphs on the
critical curve, labelled $c$ in Fig. \ref{alpha2}, and referred to as a
scaling law in Ref. \cite{Kob95}, becomes better for $d^{s}=0.5$ than for
the result shown for $d^{s}=0.6$. Let us emphasize, that the found results
are not universal features of the MCT bifurcation dynamics. The apparent
scaling gets disturbed if $d^{s}$ increases above $0.6$ and it is predicted
to be absent for $d^{s}=1.0$ as is shown in the lower panel of Fig. \ref
{alpha2}.

It was shown by Kob et al. \cite{Kob97}, that the $\alpha _{2}$ peak is
produced by clusters of particles which move faster than the ones in their
neighborhood. Only as few as $5\%$ of all particles are involved in the
formation of these clusters. It would be very surprising if the MCT, which
works with averaged quantities like $\Phi _{q}\left( t\right) $ and $\Phi
_{q}^{s}\left( t\right) $, could reproduce such subtlety of the microscopic
dynamics. This holds even more so, if the speculations on a relation of the
cited cluster dynamics with polymer dynamics \cite{Donati98} could be
substantiated. On the other hand, the so far identified features of the fast
clusters are not in an obvious contradiction to the idea that they are
representative configurations building up backflow patterns. Moreover, the
MCT equations for structural relaxation have originally been proposed and
tested against experiments for a treatment of backflow phenomena \cite
{Goetze76}. Obviously it would be helpful to carry out a first--principle
MCT calculation of $\alpha _{2}\left( t\right) $ for the cited binary
mixture along the same lines as done in this paper for the hard--sphere
system. Thereby one could clarify whether or not the qualitative agreement
between the hard--sphere--system results for $d^{s}$ about $0.6$ with the
simulation results in Ref. \cite{Kob95} is just an accident. %
\acknowledgements We cordially thank Herman Cummins, Walter Kob, and Bill
van Megen for stimulating discussions and helpful comments on our
manuscript. {Our work was supported by Verbundprojekt BMBF 03G04TUM and the
DFG grants Fu309/2--1/2.} 
\bibliographystyle{prsty}

\begin{thebibliography}{10}

\bibitem{Leutheusser84}
E. Leutheusser, Phys. Rev. A {\bf 29},  2765  (1984).

\bibitem{Bengtzelius84}
U. Bengtzelius, W. G{\"o}tze, and A. Sj{\"o}lander, J. Phys. C {\bf 17},  5915
  (1984).

\bibitem{Goetze91b}
W. G{\"o}tze,  in {\em Liquids, Freezing and Glass Transition}, edited by J.-P.
  Hansen, D. Levesque, and J. Zinn-Justin (North--Holland, Amsterdam, 1991),
  p.\ 287.

\bibitem{Goetze92}
W. G{\"o}tze and L. Sj{\"o}gren, Rep. Prog. Phys. {\bf 55},  241  (1992).

\bibitem{Mezei91}
F. Mezei, Ber. Bunsenges. Phys. Chem. {\bf 95},  1118  (1991).

\bibitem{Li92}
G. Li, W.~M. Du, X.~K. Chen, H.~Z. Cummins, and N.~J. Tao, Phys. Rev. A {\bf
  45},  3867  (1992).

\bibitem{Bartsch92}
E. Bartsch, M. Antonietti, W. Schupp, and H. Sillescu, J. Chem. Phys. {\bf 97},
   3950  (1992).

\bibitem{Megen93}
W. van Megen and S.~M. Underwood, Phys. Rev. Lett. {\bf 70},  2766  (1993).

\bibitem{Megen94b}
W. van Megen and S.~M. Underwood, Phys. Rev. E {\bf 49},  4206  (1994).

\bibitem{Baschnagel94}
J. Baschnagel, Phys. Rev. B {\bf 49},  135  (1994).

\bibitem{Kob95}
W. Kob and H.~C. Andersen, Phys. Rev. E {\bf 51},  4626  (1995).

\bibitem{Yang96}
Y. Yang, L.~J. Muller, and K.~A. Nelson, Matter. Res. Soc. Symp. Proc. {\bf
  407},  145  (1996).

\bibitem{Ma96}
J. Ma, D.~V. Bout, and M. Berg, Phys. Rev. E {\bf 54},  2786  (1996).

\bibitem{Gallo96}
P. Gallo, F. Sciortino, P. Tartaglia, and S.-H. Chen, Phys. Rev. Lett. {\bf
  76},  2730  (1996).

\bibitem{Lunkenheimer97b}
P. Lunkenheimer, A. Pimenov, and A. Loidl, Phys. Rev. Lett. {\bf 78},  2995
  (1997).

\bibitem{Cummins97}
H.~Z. Cummins, G. Li, W. Du, Y.~H. Hwang, and G.~Q. Shen, Prog. Theor. Phys.
  {\bf 126},  21  (1997).

\bibitem{Toelle97}
A. T{\"o}lle, H. Schober, J. Wuttke, and F. Fujara, Phys. Rev. E {\bf 56},  809
   (1997).

\bibitem{Franosch97}
T. Franosch, M. Fuchs, W. G{\"o}tze, M.~R. Mayr, and A.~P. Singh, Phys. Rev. E
  {\bf 55},  7153  (1997).

\bibitem{Schilling97}
R. Schilling and T. Scheidsteger, Phys. Rev. E {\bf 56},  2932  (1997).

\bibitem{Franosch97c}
T. Franosch, M. Fuchs, W. G{\"o}tze, M.~R. Mayr, and A.~P. Singh, Phys. Rev. E
  {\bf 56},  5659  (1997).

\bibitem{Petry91}
W. Petry, E. Bartsch, F. Fujara, M. Kiebel, H. Sillescu, and B. Farago, Z.
  Phys. B {\bf 83},  175  (1991).

\bibitem{Megen98}
W. van Megen, T. Mortensen, J. M{\"u}ller, and S. Williams, preprint  (1998).

\bibitem{Zorn97}
R. Zorn, Phys. Rev. B {\bf 55},  6249  (1997).

\bibitem{Kanaya97}
T. Kanaya, I. Tsukushi, and K. Kaji, Prog. Theor. Phys. Suppl. {\bf 126},  133
  (1997).

\bibitem{Sciortino96}
F. Sciortino, P. Gallo, P. Tartaglia, and S.-H. Chen, Phys. Rev. E {\bf 54},
  6331  (1996).

\bibitem{Fuchs95}
M. Fuchs, Transp. Theory Stat. Phys. {\bf 24},  855  (1995).
   
\bibitem{Megen91}
W. van Megen and P.~N. Pusey, Phys. Rev. A {\bf 43},  5429  (1991).

\bibitem{Megen91n}
W. van Megen and S.~M. Underwood and P.~N. Pusey, Phys. Rev. Lett. {\bf 67},  1586  (1991).

\bibitem{Megen93n}
W. van Megen and S.~M. Underwood, Phys. Rev. E {\bf 47},  248  (1993).

\bibitem{Megen94n}
W. van Megen and S.~M. Underwood, Phys. Rev. Lett. {\bf 72},  1773  (1994).

\bibitem{Megen95n}
W. van Megen, Transp. Theory Stat. Phys. {\bf 24},  1017  (1995).

\bibitem{Lebowitz63}
J.~L. Lebowitz, Phys. Rev. {\bf 133},  A895  (1963).

\bibitem{Goetze95b}
W. G{\"o}tze and L. Sj{\"o}gren, J. Math. Analysis and Appl. {\bf 195},  230
  (1995).

\bibitem{Franosch94}
T. Franosch and W. G{\"o}tze, J. Phys.: Condens. Matter {\bf 6},  4807  (1994).

\bibitem{Franosch98}
T. Franosch, W. G{\"o}tze, M.~R. Mayr, and A.~P. Singh, J. Non--Cryst. Solids
  (1998), in print.

\bibitem{Hansen86}
J.-P. Hansen and I.~R. McDonald, {\em Theory of Simple Liquids}, 2nd ed.
  (Academic Press, London, 1986).

\bibitem{Rahman62}
A. Rahman, K.~S. Singwi, and A. Sj{\"o}lander, Phys. Rev. {\bf 126},  986
  (1962).

\bibitem{Rahman64}
A. Rahman, Phys. Rev. A {\bf 136},  405  (1964).

\bibitem{Kob95b}
W. Kob and H.~C. Andersen, Phys. Rev. E {\bf 52},  4134  (1995).

\bibitem{Kob95c}
W. Kob and H. Andersen, Transp. Theory Stat. Phys. {\bf 24},  1179  (1995).

\bibitem{Nauroth97}
M. Nauroth and W. Kob, Phys. Rev. E {\bf 55},  657  (1997).

\bibitem{Kob97}
W. Kob, C. Donati, S.~J. Plimpton, P.~H. Poole, and S.~C. Glotzer, Phys. Rev.
  Lett. {\bf 79},  2827  (1997).

\bibitem{Donati98}
C. Donati, J.~F. Douglas, W. Kob, S.~J. Plimpton, P.~H. Poole, and S.~C.
  Glotzer, Phys. Rev. Lett. {\bf 80}, 2338  (1998).

\bibitem{Goetze76}
W. G{\"o}tze and M. L{\"u}cke, Phys. Rev. B {\bf 13},  3825  (1976).

\end{thebibliography}

\begin{figure} 
\caption[]{\label{CritAmpR1}
Critical Lamb--M\"{o}ssbauer factor $f_{q}^{sc}$ and
amplitudes $h_{q}^{s}$, $K_{q}^{s}$, $\bar{K}_{q}^{s}$, obtained from Eqs. (%
\ref{LMF}), (\ref{AMP_SISF}), and (\ref{Kbar}), for a tagged hard--sphere
particle of diameter $d^{s}=1.0$ immersed in a hard--sphere liquid. The
solid lines are the Gaussian approximations, Eqs. (\ref{GA_Crit_Amp}). The
arrows indicate the wave numbers $q_{0}=3.4$, $q_{1}=7.0$, $q_{2}=10.6$, and 
$q_{3}=17.4$. The dotted line in the upper panel is one tenth of the
structure factor at the critical point (compare Fig. $3$ in Ref. \cite
{Franosch97}). Here and in the following figures the diameter $d$ of the
host particles is chosen as the unit of length.
} 
\end{figure} 
\begin{figure} 
\caption[]{\label{CritAmpR06}
Same as Fig. \ref{CritAmpR1} for a tagged particle
of diameter $d^{s}=0.6$.

} \end{figure} 
\begin{figure} \caption[]{  \label{phis of t}Self--intermediate scattering function $\Phi
_{1}^{s}(t)$, calculated from Eqs. (\ref{MCE_SISF}) and (\ref{DMCF_SISF}),
of the HSS ($d^{s}=1.0$) for wave number $q_{1}=7.0$ at various reduced
packing fractions $\varepsilon =\left( \varphi -\varphi _{c}\right) /\varphi
_{c}=\pm 10^{-n/3}$, where $n=0,\ldots ,14$. The thick curve labelled by $c$%
, shows the dynamics at the critical packing fraction $\varphi
_{c}=0.515912\ldots $. The uppermost curve refers to the packing fraction $%
\varphi =0.6$. The arrow marks the time $t_{0}=0.425$. Here and in some of
the following figures the full dot and square mark the times $t_{\sigma }$
and $t_{\sigma }^{\prime }$, respectively, for $\varepsilon =-0.001$. On the
glass side ($\varepsilon >0$) the curves for $n=13,14$ are omitted since
they would be barely distinguishable from curve $c$. The $n=0,1,2$ glass
curves are omitted, since the Percus--Yevick approximation produces negative
values for the pair distribution function for such high densities.
Here and in the following figures the unit of time is chosen such that the
short--time diffusion coefficient of the host particles reads $D_0=1/160$.
} \end{figure} 
\begin{figure} \caption[]{  \label{red_theo}Upper panel: Rescaled self--intermediate scattering
function $\hat{\Phi}_{q}^{s}\left( t\right) =\left( \Phi _{q}^{s}\left(
t\right) -f_{q}^{sc}\right) /h_{q}^{s}$ (solid lines) of the hard--sphere
system ($d^{s}=1.0$) for four different wave numbers at reduced packing
fraction $\varepsilon =\left( \varphi -\varphi _{c}\right) /\varphi _{c}=\pm
10^{-n/3}$ with $n=9$, showing collapse onto the $\beta $ correlator $G$
[dashed lines, Eq. (\ref{beta_scaling})] in the $\beta $--relaxation region.
On the extreme left and right small numbers near each curve indicate the
wave numbers $q_{0}$, $q_{1}$, $q_{2}$ and $q_{3}$, marked in Fig. \ref
{CritAmpR1}. Lower panel: Rescaled intermediate scattering function (ISF)
and self--intermediate scattering function (SISF) for wave number $q_{2}=10.6
$ at $n=10$, compared to the $\beta $ correlator $G$. The glass curves ($%
\varepsilon >0$) are shifted downward by $1$ to avoid overcrowding. The open
symbols mark the boundaries of the $\beta $--relaxation region as defined at
the end of Sect. \ref{BETA_SISF}. Some symbols are missing because they are
either outside the display range (the long--time part of the liquid curve
for $q_{1}$) or the range of validity is non--existent (glass curves for $%
q_{0}$, $q_{1}$ and $q_{3}$).

} \end{figure} \begin{figure} \caption[]{  \label{ascale_phist}$\alpha $--scaling plot of the self--intermediate
scattering function: $\Phi _{1}^{s}\left( t\right) $, taken from Fig. \ref
{phis of t}, for reduced packing fractions $\varepsilon =\left( \varphi
-\varphi _{c}\right) /\varphi _{c}=-10^{-n/3}$, $n=1,3,5,7,9$ versus
rescaled time $\tilde{t}=t/t_{\sigma }^{\prime }$. The thick solid line is
the $\alpha $ master function $\tilde{\Phi}_{q}^{s}\left( \tilde{t}\right) $%
. The diamonds mark the early--time bound for the $\alpha $--scaling regime
as defined at the end of Sect. \ref{BETA_SISF}. The dotted line indicates $%
\tilde{\Phi}_{q}^{s}\left( \tilde{t}\right) -\sigma h_{q}B_{1}\tilde{t}^{b}$%
, which includes the small--$\tilde{t}$ correction according to Eq. (\ref
{alpha_corr}).

} \end{figure} \begin{figure} \caption[]{  \label{r2}Mean--square displacement $\delta r^{2}\left( t\right) $,
obtained from Eqs. (\ref{MCE_MSD}) and (\ref{DMCF_MSD}), for the
hard--sphere system ($d^{s}=1.0$) at various packing fractions. The solid
curves are labelled as in Fig. \ref{phis of t}. The straight dotted line
with unit slope indicates the long--time--diffusion asymptote ($6D^{s}t$)
for the $n=9$ liquid curve.

} \end{figure} \begin{figure} \caption[]{  \label{corr_r2}Mean--square--displacement curves $\delta r^{2}\left(
t\right) $ (full lines) taken for $n=9$ from Fig. \ref{r2}. The dashed and
the dot--dashed lines are respectively the leading and the next--to--leading
order of the asymptotic expansion (\ref{AE_MSD}) with $r_{sc}^{2}=5.57\cdot
10^{-3}$, $h_{MSD}=0.0116$, $K_{MSD}=-1.23$ and $\hat{K}_{MSD}=3.33$,
calculated from Eqs. (\ref{AMP_MSD}). The open symbols indicate the range of
validity of the first ($\Diamond $) and second ($\bigcirc $) order formulae
as defined at the end of Sect. \ref{BETA_SISF}. On the liquid side the
second--order range of validity extends beyond the displayed range. On the
glass side the first--order range of validity does not exist. The dotted
straight line of slope unity for small times indicates the short--time
diffusion asymptote ($6D_{0}^{s}t$).

} \end{figure} \begin{figure} \caption[]{  \label{Diff_Coeff}Self--diffusion coefficients $D^{s}$ for tagged
particles of diameters $d^{s}=1.0$ and $d^{s}=0.6$ and $\alpha $--relaxation
rate $1/\tau _{1}^{\alpha s}$ versus reduced packing fraction $\varepsilon
=\left( \varphi -\varphi _{c}\right) /\varphi _{c}$. The solid lines are the
power--law asymptotes ($\Gamma _{A}\left| \varepsilon \right| ^{\gamma }$, $%
\gamma =2.46$). The $\alpha $--scaling time $\tau _{1}^{\alpha s}$ is
defined as the time when the tagged--particle density correlator $\Phi
_{1}^{s}(t)$, shown in Fig. \ref{phis of t}, has relaxed to half of its
critical plateau value $f_{1}^{sc}=0.760$.

} \end{figure} \begin{figure} \caption[]{  \label{ascale_r2}$\alpha $--scaling plot of the mean--square
displacement: $\delta r^{2}\left( t\right) $, taken from Fig. \ref{r2}, for
reduced packing fractions $\varepsilon =\left( \varphi -\varphi _{c}\right)
/\varphi _{c}=-10^{-n/3}$, $n=1,3,5,7,9$ versus rescaled time $\tilde{t}%
=t/t_{\sigma }^{\prime }$. The thick solid line is the $\alpha $ master
function $\delta \tilde{r}^{2}\left( \tilde{t}\right) $ calculated from Eq. (%
\ref{alpha_MSD}). The diamonds mark the early--time bound for the $\alpha $%
--scaling regime as defined at the end of Sect. \ref{BETA_SISF}. The dashed
line shows the master function corrected according to the analogue of Eq. (%
\ref{alpha_corr}). Leaving aside the $\tilde{t}$--independent bracket term
one gets the dotted curves.

} \end{figure} \begin{figure} \caption[]{  \label{Gauss_Approx}Test of the Gaussian approximation (\ref{EGA})
(dashed lines) and of the cumulant expansion (\ref{cumulant}) (dot--dashed
lines) for the self--intermediate scattering function of the HSS (solid
lines) at packing fraction $\varphi =\varphi _{c}\left( 1\pm
10^{-7/3}\right) $ for the wave numbers $q_{0}$, $q_{1}$, $q_{2}$ and $q_{3}$%
, introduced in Fig. \ref{CritAmpR1}.

} \end{figure} \begin{figure} \caption[]{  \label{alpha2}Non--Gaussian parameter $\alpha _{2}\left( t\right) $,
obtained from Eq. (\ref{MCE_NGP}), for tagged particles of diameters $%
d^{s}=1.0$ and $d^{s}=0.6$. The labelling is done as in Fig. \ref{phis of t}.

} \end{figure} \begin{figure} \caption[]{  \label{corr_a2}$\beta $--relaxation behavior of the non--Gaussian
parameter $\alpha _{2}\left( t\right) $ (full lines) taken for $n=9$ from
Fig. \ref{alpha2}. The dashed and the dot--dashed lines are respectively the
leading and the next--to--leading--order asymptotic expansion (\ref{AE_NGP}%
). The critical amplitudes ($d^{s}=1.0$: $f_{NGP}=-0.0445$, $h_{NGP}=-1.12$, 
$K_{NGP}=0.543$, $\hat{K}_{NGP}=-2.98$; $d^{s}=0.6$: $f_{NGP}=0.891$, $%
h_{NGP}=-1.37$, $K_{NGP}=3.61$, $\hat{K}_{NGP}=-9.85$) were calculated from
Eqs. (\ref{AMP_A2}). The symbols indicate the range of validity of the
leading ($\Diamond $) and the next--to--leading ($\bigcirc $) order as
defined at the end of Sect. \ref{BETA_SISF}---the first--order range of
validity indicated in the lower panel is due to an ''accidental''
intersection of curves.

} \end{figure} \begin{figure} \caption[]{  \label{ascale_a2}$\alpha $--scaling plot of the non--Gaussian
parameter: $\alpha _{2}\left( t\right) $, taken from Fig. \ref{alpha2}, for
reduced packing fractions $\varepsilon =\left( \varphi -\varphi _{c}\right)
/\varphi _{c}=\pm 10^{-n/3}$, $n=1,3,5,7,9$ versus rescaled time $\tilde{t}%
=t/t_{\sigma }^{\prime }$. The thick solid line is the $\alpha $ master
function $\tilde{\alpha}_{2}\left( \tilde{t}\right) $. The dashed line shows
the master function corrected according to the analogue of Eq. (\ref
{alpha_corr}). Leaving aside the $\tilde{t}$--independent bracket term one
gets the dotted curves.
} \end{figure} 
\end{mathletters}

\end{document}